\documentclass[aps,preprintnumbers,amsmath,amssymb,endfloats]{revtex4} 
\usepackage{dcolumn}
\usepackage{bm}

\pdfoutput=1

   \ifx\pdfoutput\undefined
   \usepackage[dvips]{graphicx}
   \else
   \usepackage[pdftex]{graphicx}
   \fi
   \usepackage{epstopdf}
   \DeclareGraphicsExtensions{.pdf,.gif,.jpg,.eps,.png}
  
   \usepackage{amsmath}
  \usepackage{amssymb}

\begin{document}

\preprint{{\it submitted to The Physics of Plasmas}}

\title{The effects of strong temperature anisotropy on the kinetic structure
of collisionless slow shocks and reconnection exhausts. Part I: PIC
simulations}
 
\author{Yi-Hsin~Liu}
\affiliation{University of Maryland, College Park, MD 20742}
\author{J.~F.~Drake}
\affiliation{University of Maryland, College Park, MD 20742}	
\author{M.~Swisdak}
\affiliation{University of Maryland, College Park, MD 20742}

\date{\today}

\begin{abstract}
A 2-D Riemann problem is designed to study the development and
dynamics of the slow shocks that are thought to form at the boundaries
of reconnection exhausts. Simulations are carried out for varying
ratios of normal magnetic field to the transverse upstream magnetic field ({\it
i.e.}, propagation angle with respect to the upstream magnetic field).
When the angle is sufficiently oblique, the simulations reveal a large
firehose-sense ($P_\|>P_\bot$) temperature anisotropy in the downstream region,
accompanied by a transition from a coplanar slow shock to a
non-coplanar rotational mode.
In the downstream region the firehose stability parameter 
$\varepsilon=1-\mu_0(P_\|-P_\perp)/ B^2$ tends to lock in to 0.25. 
This balance arises from the competition between counterstreaming ions, 
which drives $\varepsilon$ down, and the scattering due to ion inertial scale waves, 
which are driven unstable by the downstream rotational wave.
At very oblique propagating angles,
2-D turbulence also develops in the downstream region.
\end{abstract}

\maketitle

\section{Introduction}

Following the publication of the MHD reconnection scenario of Sweet
and Parker \cite{sweet58a, parker57a}, Petschek
\cite{petschek64a} noted that a pair of back-to-back slow shocks
bounding the reconnection outflow could significantly raise the
efficiency of the process by acting as a transition between the
inflowing and reconnected outflowing plasma.  Ion heating by these
Petschek-reconnection-associated slow shocks is one of the mechanisms
that has been proposed for solar flares
\cite{tsuneta96a, longcope09a} and the solar wind.  However, although in-situ
observations of slow shocks in the solar wind exist
\cite{feldman84a,feldman87a, smith84a,oieroset00a,hoshino00a,walthour94a}, they
are relatively rare, suggesting that the MHD picture of Petschek may
not tell the complete story.  Unanswered questions remain concerning
the kinetic structure of such shocks in a collisionless plasma and the
associated mechanisms leading to particle heating.  In previous work,
kinetic slow shocks were studied numerically in hybrid codes by
initializing the system with the slow shock jumps predicted by
MHD \cite{swift83a} or, later, by the piston \cite{winske85a} and
flow-flow methods \cite{omidi92a}. Some of the main focuses of these
works were the backstreaming beam-driven electromagnetic ion-ion
cyclotron instability (EMIIC) \cite{winske85a,omidi92a}, which has
been suggested to be the cause of the nonsteady behavior of slow
shocks, and the formation and damping of downstream large amplitude
dispersive wavetrains \cite{swift83a,lin93a}. Recently, the
dissipation due to electrons and beam-excited kinetic Alfv\'en waves
was studied by the piston method in particle-in-cell (PIC) simulations
\cite{yin05a,yin07b}.

In PIC simulations of reconnection, the plasma downstream of the X-line
exhibits large firehose-sense ($P_\|>P_\bot$) temperature anisotropies, as shown in
Fig.~\ref{LO_reconn}. (This simulation was discussed previously
\cite{drake09a}). Panel (a) shows the out-of-plane electron current,
with the X-point at $(x/d_i, z/d_i)\sim(0, 20)$, where $d_i$ is the
ion inertial scale. In panels (b) and (c), we see turbulence in the
$B_x$ component correlated with the unstable firehose region; a cut of
the firehose stability parameter $\varepsilon=1-\mu_0(P_\|-P_\perp)/
B^2$ at $z=-35 d_i$ is shown in (d). The firehose-sense temperature
anisotropy is notable since in-situ observations of the solar wind
clearly show that the proton temperature anisotropy is bounded by the
marginal firehose and mirror mode stability boundaries
\cite{bale09a}. Hence, it is of interest to more closely study the
temperature anisotropy distribution across the reconnection exhaust
far downstream from the reconnection site, in order to understand the
effect of the self-generated temperature anisotropy on the propagation
and steepening of the slow shock.

The Petschek theory of reconnection \cite{petschek64a} predicts that
the reconnection exhaust will be bounded by a back-to-back pair of
standing switch-off slow shocks. However, no clear signature of
Petscheck shocks has been seen in PIC reconnection simulations, such
as in Fig. \ref{LO_reconn} \cite{drake09a}, or hybrid simulations
\cite{lottermoser98a, nakamura98a}.  This may be due to the relatively
small domain sizes in the shock normal direction ($\hat{{\bf e}}_x$ in
Fig.~\ref{LO_reconn}), although other reasons have been discussed in
the context of large-scale hybrid reconnection simulations
\cite{lin96a,lottermoser98a}. To address this issue, we perform 2-D
PIC simulations that extend the simulation size in the normal
direction to $\sim 800 d_i$ by ignoring the X-line and instead
examining the conceptually simpler Riemann problem.  In the Riemann
formulation, waves propagate away from an interface of two different
uniform states, such as the two sides of a reconnection symmetry
line. This set-up more closely resembles the reconnection outflow
exhaust than that produced by other methods of generating
shocks. Similar Riemann problems have been carried out in 1-D
resistive-MHD
\cite{lin93a}, including analyses of asymmetric states and the effect
of a guide field, and in 1-D hybrid simulations
\cite{lin96a}. 2-D Riemann problems have also been carried out in
hybrid simulations \cite{scholer98a} where the 2-D downstream
turbulence appeared to diminish the downstream wavetrains associated
with switch-off slow shocks. Upstream perpendicular heating by EMIIC
and the subsequent excitation of Alfv\'en/ion cyclotron waves have
also been studied in similar 2-D Riemann
problems \cite{cremer99a,cremer00a}.

In Sec.~II of this paper we introduce our simulation model and the
initial set-up of the Riemann problem.  In Sec.~III we discuss the
general profiles of a run with $\theta_{BN}=75^\circ$ (the angle
between the upstream magnetic field and the normal direction (${\bf
e}_x$)).  Section IV points out that the counterstreaming ions drive
$\varepsilon$ down (increase the firehose-sense temperature
anisotropy) in the downstream region.  In Sec.~V we show that a more
oblique angle results in a lower $\varepsilon$ at downstream
region. The structure of the magnetic field performs a transition from
a coplanar decrease to a non-coplanar rotation at $\varepsilon \sim
0.25$, which differs from the traditional slow shock transition with
dispersive wavetrains.  In Sec.~VI the stability of the downstream
rotational wave is studied with numerical experiments. The tendency
for a spatially modulated rotational wave to radiate $d_i$-scale waves
is identified. The resulting $d_i$-scale waves counter-balance the
$\varepsilon$ decrease driven by the counterstreaming ions.  Finally,
we summarize the results and discuss potential implications in
Sec.~VII.

\section{simulation models and details}

Our PIC simulations use a narrow computational domain, $l_z\times l_x
= 1.6 d_i \times 1638.4 d_i$ to capture the nonlinear wave propagation
(mainly the slow shock pair in Petscheck's reconnection model) far
downstream from the reconnection site. The strategy is to use time as a
proxy for space in order to reduce the computational burden.  The simulations
presented here are two-dimensional, {\it i.e.}, $\partial/\partial y =
0$, and periodic in the $z-x$ plane. The initial equilibrium consists
of a double Harris-like current sheet (although we only focus on a single
current sheet) superimposed on an ambient population of uniform
density $n_a$:
\begin{equation}
 B_z=B_{z,a}\tanh(x/w_i); \qquad n_{p,e}=n_h\mbox{sech}^2(x/w_i)+n_a, 
 \label{inhomo_pic1}
\end{equation}
where $B_{z,a}, n_h, n_a $ are constants, the subscript ``a'' stands
for the asymptotic (far upstream) values, ``h'' stands for Harris and
$w_i$ is the initial half-width of the current sheet. We initialize
both the Harris plasma and background plasma with an isotropic
Maxwellian distribution. Unlike the initial set-up for reconnection,
we begin with a constant normal field $B_x$. Although the initial
total pressure is balanced, the existence of $B_x$ causes a tension
force that drives wave propagation away from the current sheet in the
$x$-direction. The $B_{x,a}/B_{z,a}=0.1$ case ({\it i.e.,} $\theta_{BN}
\sim 83^\circ$ with $\cos\theta_{BN}\equiv B_{x,a}/B_{a}$) 
corresponds to a reconnection exhaust with a normalized reconnection
rate of 0.1.

This simulation box can be pictured as existing in the reconnection
outflow frame. The distance from the reconnection site is estimated as
$C_{Az} \times \text{time}$, where $C_{Az}\equiv B_{z,a}/\sqrt{\mu_0
m_i n_a}$ is the Alfv\'en speed based on the reversed component of the
field. When $l_z$ is small our simulations are essentially 1-D,
although we do perform runs with larger $l_z$ to investigate the
possibility of developing 2-D turbulence.  Waves ({\it e.g.,} fast,
intermediate, and slow modes in the fluid model) will propagate away
from the central discontinuity at their characteristic speeds, and can
steepen into shocks, spread into rarefactions, or maintain their
initial shapes based on their own nonlinearities.

In our particle-in-cell code p3d \cite{zeiler02a}, the electromagnetic
fields are defined on gridpoints and advanced in time with an explicit
trapezoidal-leapfrog method using second-order spatial derivatives.
The Lorentz equation of motion for each particle is evolved by a Boris
algorithm where the velocity $\mathbf{v}$ is accelerated by
$\mathbf{E}$ for half a timestep, rotated by $\mathbf{B}$, and
accelerated by $\mathbf{E}$ for the final half timestep.  To ensure
that $\bm{\nabla \cdot} \mathbf{E} =\rho/\epsilon_0 $ a correction
electric field is calculated by inverting Poisson's equation with a
multigrid algorithm.

The magnetic field is normalized to the asymptotic magnetic field
$B_a$, the density to the asymptotic density $n_a$, velocities to the
Alfv\'en speed $C_A\equiv B_a/\sqrt{\mu_0 m_i n_a}$, lengths to the
ion inertial length $d_i \equiv \sqrt{m_i/\mu_0 n_a e^2}$, times to
the inverse ion cyclotron frequency $\Omega_{ci}^{-1}\equiv m_i/B_ae$,
and temperatures to $m_i C_{A}^2$.  Other important parameters are
$m_i/m_e= 25$, $c/C_A=15$, $n_a=1$, $n_h=1.5$, $B_a=1$, and the
asymptotic value of initial $T_{i,e}=0.1$, which implies that
$\beta_a=0.4$.  The initial electron temperature is uniform, while the
ion temperature varies so as to ensure pressure balance in the
$x$-direction.  We take the time step $\Delta t=0.0025$ and grid size
$\Delta=0.025$. We usually take $w_i=d_i$ since the thickness of the
dissipation region during reconnection is on the $d_i$ scale. There
are $\sim 4\times 10^8$ particles in a single run. Table I gives
further details of the various runs. \newline

\begin{table}
\caption{}
\begin{center}
\begin{tabular}{cccccccc}
\tableline\tableline
Run&$\theta_{BN}$ & $w_i$ & $B_g$\footnotemark[1] & Domain Size ($l_z \times l_x$) & Gridpoints  & $\rightarrow$\footnotemark[2] 2-D turbulence\\
\tableline
${\bf a}$& $30^\circ$ & 1 & 0 &$1.6 \times 1638.4$ & $64 \times 65536$ & x\\
${\bf b}$& $45^\circ$ & 1 & 0 &$1.6 \times 1638.4$ & $64 \times 65536$ & x\\
${\bf c}$& $52^\circ$ & 1 & 0 &$1.6 \times 1638.4$ & $64 \times 65536$ & x\\
${\bf d}$& $60^\circ$ & 1 & 0 &$1.6 \times 1638.4$ & $64 \times 65536$ & x\\
${\bf e}$& $60^\circ$ & 1 & 0 &$6.4 \times 819.2$ & $256 \times 32768$ & x\\ 
${\bf f}$& $75^\circ$ & 1 & 0 &$1.6 \times 1638.4$ & $64 \times 65536$ & x\\
${\bf g}$& $75^\circ$ & 10 & 0 & $1.6 \times 1638.4$ & $64 \times 65536$ &x\\
${\bf h}$& $75^\circ$ & 1 & 0.2 & $1.6 \times 1638.4$ & $64 \times 65536$ &x\\
${\bf i}$& $75^\circ$ & 1 & 0 &$6.4 \times 1638.4$ & $256 \times 65536$ & $\surd$\\
${\bf j}$& $83^\circ$ & 1 & 0 &$1.6 \times 819.2$ & $64 \times 32768$ & x\\
${\bf k}$& $83^\circ$ & 1 & 0 &$6.4 \times 819.2$ & $256 \times 32768$ & $\surd$\\
\tableline
\end{tabular}
\footnotetext[1]{ $B_g$ is an initial uniform guide field in the y-direction}
\footnotetext[2]{ ``$\rightarrow$" means ``resulting in".}
\end{center}
\end{table}

Runs ${\bf a}$, ${\bf b}$, ${\bf c}$, ${\bf d}$, ${\bf f}$, ${\bf g}$,
${\bf h}$ and ${\bf k}$ (all have $l_z=1.6 d_i $, except the
$\theta_{BN}=83^\circ$ case) will be further discussed in this
work. Even though the $75^\circ$ case with a larger $l_z=6.4 d_i$ (Run
${\bf i}$) shows downstream 2-D turbulence, the evolution is quite
similar to the narrow Run ${\bf f}$.  As can be seen in TABLE I, 2-D
turbulence tends to occur for oblique ($\theta_{BN}> 75^\circ$) cases
when $l_z$ is large enough. Curves plotted throughout the rest of this
paper are quantities averaged in the $z$-direction, since most runs
discussed here do not have any significant variation in the
$z$-direction.  The 2-D turbulence of $\theta_{BN}=83^\circ$(Run ${\bf
k}$) will also be discussed.

\section{General features of the $75^\circ$ case}

\begin{figure} 
  \includegraphics[width=13cm]{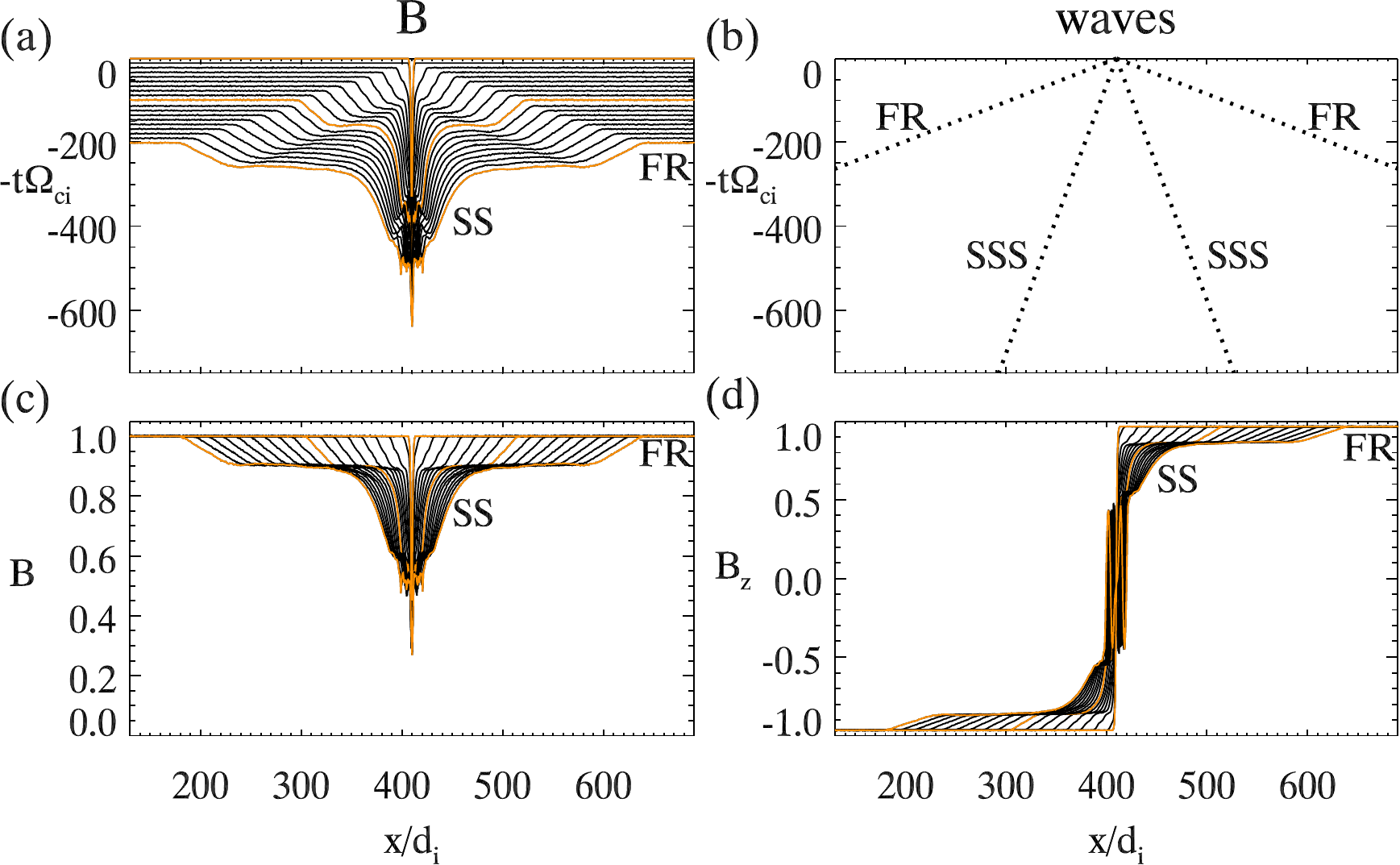} 
\caption{The evolution of a system with $\theta_{BN}=75^\circ$ (Run ${\bf f}$). Panel (a): The
evolution of $B$ from time $0-200/\Omega_{ci}$. A pair of fast
rarefactions (FR) propagate out from the symmetry line, followed by a
pair of slow shocks (SS). Each curve has been shifted so that it
intersects the vertical axis at the given time.  The time between the
yellow curves is $100/\Omega_{ci}$; Panel (b): The predicted FR and
switch-off slow shock (SSS) from ideal MHD theory; Panel (c): The same
as (a) but with the vertical axis measuring $B$; Panel (d): The
evolution of $B_z$ from time $0 - 200/\Omega_{ci}$.}  \label{LO_stack}
\end{figure}

A representative case of $\theta_{BN}=75^\circ, w_i=1d_i, B_g=0$
(Run ${\bf f}$; hereafter referred to as the ``$75^\circ$ run'') at time
$200/\Omega_{ci}$ is documented in Fig.~\ref{LO_stack}. As soon as the
simulation begins, a pair of fast rarefaction waves propagate out
from the discontinuity with speed $\sim 1.1 C_A$, while the slow
shocks with their downstream rotational waves have speed $\sim 0.15
C_A$.  As in ideal MHD with a symmetric initial condition and zero guide
field, a pair of switch-off slow shocks are expected to follow the
fast rarefaction waves, as shown in Fig. \ref{LO_stack}(b)
\cite{lin93a}. A switch-off slow shock ({\it i.e.}, the strongest slow
shock, whose tangential magnetic field vanishes downstream of the
shock) propagates at the upstream intermediate speed, while the
downstream linear slow mode speed equals the downstream linear
intermediate speed. The linear slow mode and intermediate mode are
known to become degenerate at parallel propagation when the plasma
$\beta$ (plasma thermal pressure/magnetic pressure) exceeds $1$.  The
downstream rotational waves are often identified as dispersive
wavetrains. The essential physics of this wavetrain can be described
by a two-fluid model \cite{coroniti71a}, or by Hall-MHD.

\begin{figure}
  \includegraphics[width=12cm]{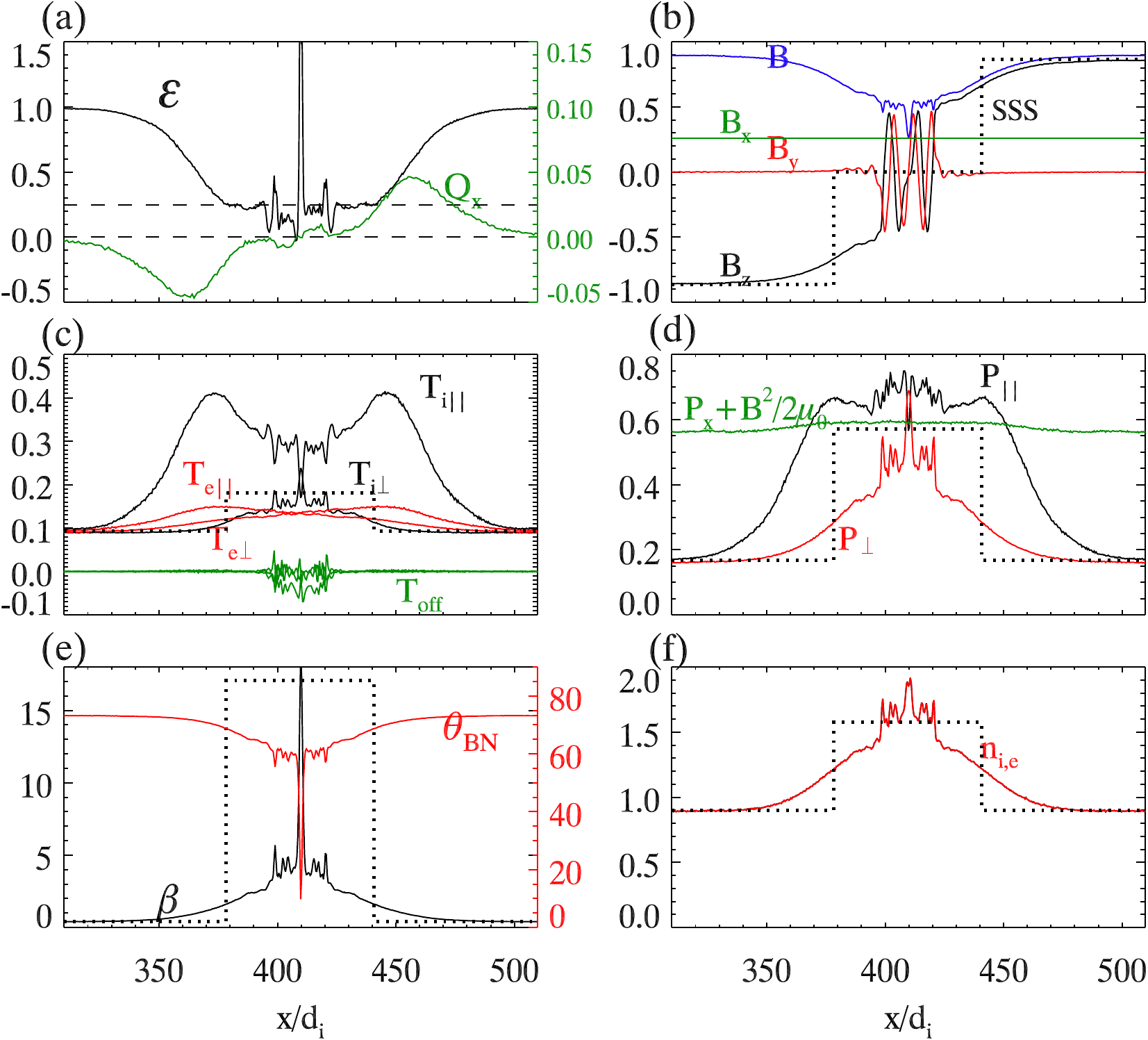} 
\caption{Parameters from the run with $\theta_{BN}=75^\circ$ (Run ${\bf f}$) at time
$200/\Omega_{ci}$. Panel (a): Temperature anisotropy $\varepsilon$ and
x-direction heat flux $Q_x$; Panel (b): Magnetic field components;
Panel (c): Parallel and perpendicular temperatures (the off-diagonal
components $T_{ixy}, T_{ixz}, T_{iyz}$ are plotted together in green,
denoted as $T_{\text{off}}$, and are small) ; Panel (d): Total plasma
pressure components and $P_x+B^2/2\mu_0$. Panel (e): The plasma
$\beta$ and local $\theta_{BN}=\cos^{-1}(B_x/B)$; Panel (f): Plasma
density. The dotted curves in each panel are the predicted magnitude
and position of the switch-off slow shocks (SSS) from isotropic MHD
for $B_z$ in (b), T in (c), P in (d), $\beta$ in (e), and n in (f).}
\label{LO_75d_200} 
\end{figure}

Further details of the slow shock pair are shown in
Fig.~\ref{LO_75d_200}. The most significant feature differing from the
ideal MHD model is the presence of a large downstream temperature
anisotropy $\varepsilon \equiv 1-\mu_0(P_\|-P_\perp)/B^2 $ shown in
panel (a). The corresponding magnetic field structure is shown in
panel (b), where the left-hand-polarized rotational wave is clearly
seen (the polarization will be discussed further in the hodograms of
Fig.~\ref{LO_degrees}). The value of $\varepsilon$ drops from $1.0$
upstream of the slow shock to $\sim0.25$ around the nearly
constant-magnitude rotational waves found downstream. The anisotropy
factor $\varepsilon$ affects the strength of the tension force, which
in the fluid theory is proportional to $\varepsilon ( {\bf B}\cdot
\nabla) {\bf B}/\mu_0$. When $\varepsilon$ is positive, the magnetic
field has a restoring tension force, while a negative value makes the
tension force operate in the opposite way, driving the firehose
instability. Viewed another way, the phase speed of an intermediate
mode is $C_I \equiv \sqrt{\varepsilon} C_A
\cos(\theta_{BN})$. Therefore, as $\varepsilon$ drops the intermediate
mode becomes slower, or even stops propagating, going firehose
unstable for $\varepsilon<0$. The $x$-direction heat flux $Q_x
\equiv \int{d^3v (\frac{1}{2}m \delta v^2 \delta v_x)f}$, where
$\delta {\bf v}\equiv {\bf v}-\langle{\bf v}\rangle$, is also
documented in (a). The heat flux peaks inside the transition from
upstream of the slow shock to the downstream rotational waves, and
then becomes negligible.  This fact is used in the follow-up paper
\cite{yhliu11b}.
In panel (c), the parallel ion temperature increases sharply in the weak field region, while the perpendicular ion
temperature is nearly constant. The electrons are nearly isotropic
across the shock. In panel (d), the parallel plasma pressure and
perpendicular plasma pressure are shown, and the nearly constant
normal direction pressure balance $P_x+B^2/2\mu_0$ indicates the
absence of fast modes in the reversal region.
Panel (e) shows the associated variations in $\beta$ and the local
$\theta_{BN}$ and panel (f) documents the density profiles.  In order
to pin down important kinetic effects not included in the MHD model
(such as the large temperature anisotropy), the black dotted curves in
each panel show the predicted jumps and positions of slow shocks in
the ideal MHD version of this global Riemann problem \cite{lin93a}.


Within the MHD predicted switch-off slow shock (SSS) jump shown in
panel (b), a coplanar transition decreases the SSS upstream $B_z$ to
$\sim 0.5$.  After this, the magnetic field structure rotates in the
non-coplanar direction and exhibits nearly constant $|B|$ inside the
downstream rotational waves. The amplitude eventually drops to the
value of $B_x$ (i.e., the tangential magnetic field vanishes) in the
center, as the symmetry of the initial conditions demands. The
coplanar transition is recognizable as a slow shock transition, where
the major enhancements of the temperatures, pressures, densities and
decrease in $|B|$ occur. A similar step-like decrease in $|B|$ due to
a slow shock was also noted in the downstream of a large-scale hybrid
reconnection simulation \cite{lottermoser98a}.  The constancy of the
total magnetic field inside the downstream rotational wave suggests an
intermediate-wave-like behavior.

\section{The source of temperature anisotropy: Alfv\'enic counter-streaming ions}\

Using the Walen relation ${\bf V}_{t,d}-{\bf V}_{t,u} = \pm
\sqrt{\rho_u \varepsilon_u/\mu_0}({\bf B}_{t,d}/\rho_d-{\bf
B}_{t,u}/\rho_u)$ for switch-off slow shocks or rotational
discontinuities (``t'' for tangential, ``u'' for upstream and ``d''
for downstream), an outflow in the $z$-direction with Alfv\'enic
velocity, $B_{z,u}/\sqrt{\mu_0 \rho_u}\sim C_{Az}\equiv
B_{z,a}/\sqrt{\mu_0 m_i n_a}$ is predicted \cite{sonnerup81b}. The
energy source of the downstream outflow is the difference in the
tangential component of the magnetic field across the
discontinuity. If we jump to the outflow frame (the de Hoffmann-Teller
frame), there will be inflowing Alfv\'enic streaming ion beams from
both discontinuities along the downstream magnetic field as observed
by Gosling in the solar wind
\cite{gosling05a} and in kinetic reconnection simulations
\cite{krauss-varban95a,lottermoser98a,hoshino98b,nakamura98a, drake09a}. 
These counter-streaming ions cause an
enhancement in the downstream parallel ion temperature and, therefore,
the temperature anisotropy.

\begin{figure} 
   \includegraphics[width=12cm]{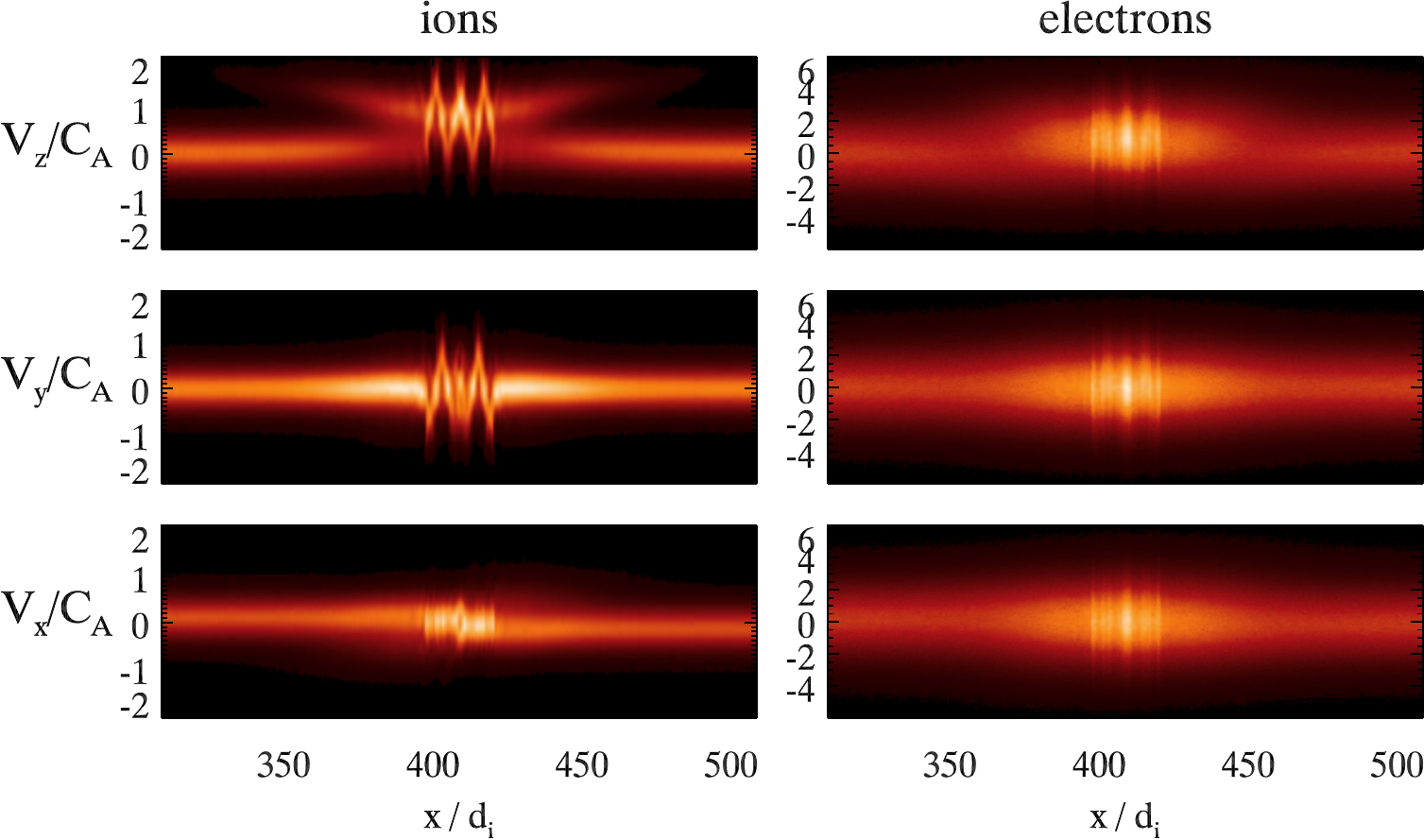} 
\caption{The phase space of the run with $\theta_{BN}=75^\circ$ (Run ${\bf f}$) at time
$200/\Omega_{ci}$. From top to bottom the left column shows the ion
distribution in: $V_z-x$ space, where the backstreaming ions from the
discontinuities are clearly seen; $V_y-x$ space; $V_x-x$ space. The
right column is the electron distribution in $V_z-x$ space, $V_y-x$
space and $V_x-x$ space.} \label{LO_phase} \end{figure}

In the phase space of the $75^\circ$ case at time $200/\Omega_{ci}$
(see Fig.~\ref{LO_phase}) a signature of the counter-streaming beams
is not obvious in the downstream region (since the ion distribution
does not peak in the upper and lower parts of a single wave
oscillation), perhaps because of the large-amplitude rotational wave.
However, Alfv\'enic backstreaming ions
in the $z$-direction (close to the parallel direction in the upstream
region) are observed in Fig.~\ref{LO_phase}. The time-of-flight
effect (faster ions escape farther upstream) slowly broadens the
transition region of slow shocks with time.
The nearly uniform electron distribution in all directions is due to
their high thermal conductivity and much lighter mass compared to the
ions.

\section{Temperature anisotropy vs. propagation angles}\label{ta_section}
  
\begin{figure} \includegraphics[width=15cm]{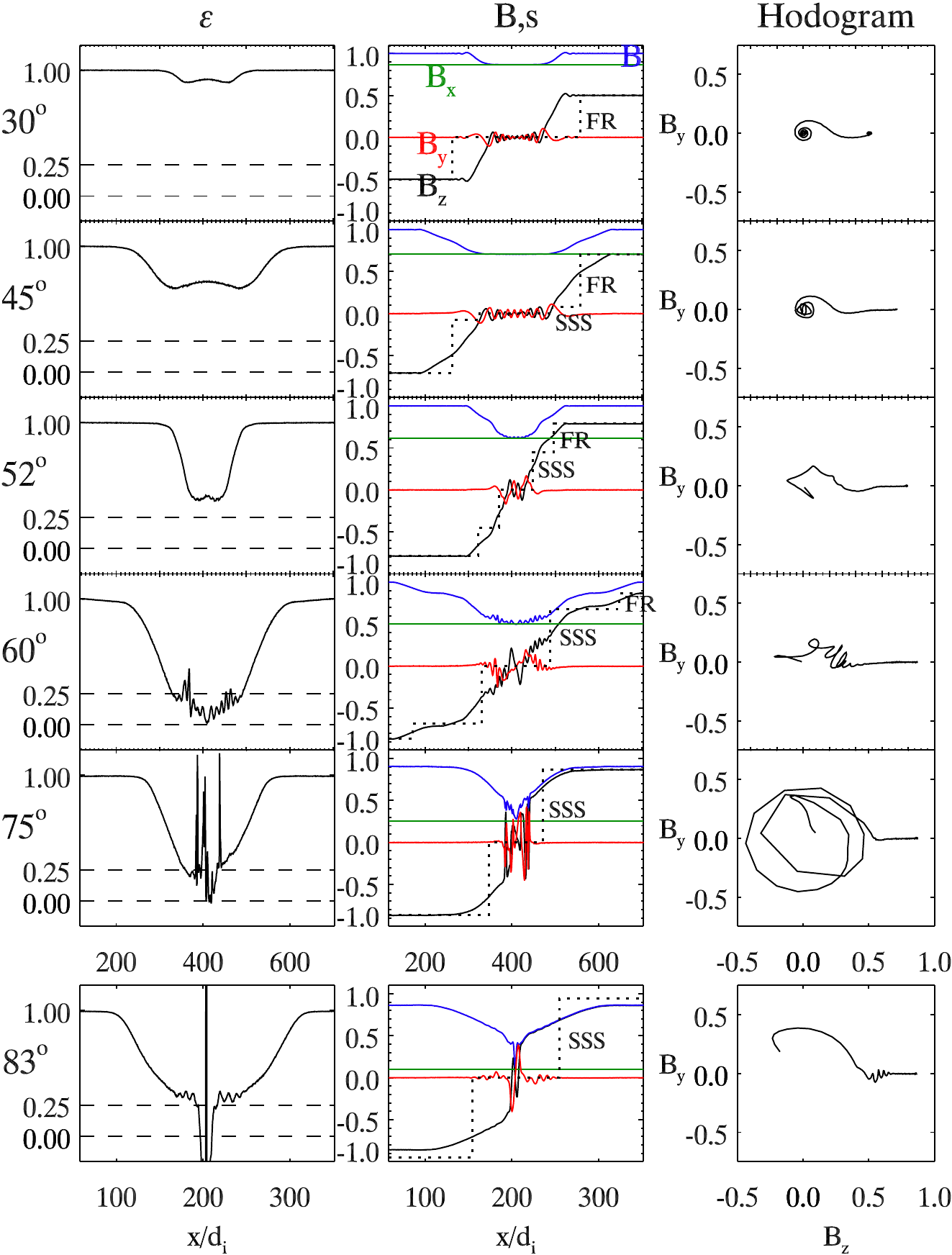} 
\caption{ From top to bottom are runs with $\theta_{BN}=
30^\circ$ (Run ${\bf a}$) at $100/\Omega_{ci}$, $45^\circ$ (Run ${\bf b}$) at $200/\Omega_{ci}$,
$52^\circ$ (Run ${\bf c}$) at $100/\Omega_{ci}$, $60^\circ$ (Run ${\bf d}$) at $250/\Omega_{ci}$,
$75^\circ$ (Run ${\bf f}$) at $400/\Omega_{ci}$, and $83^\circ$ (Run ${\bf k}$) at
$700/\Omega_{ci}$. The first column shows the temperature anisotropy,
and the second column the magnetic field components as a function of
$x$. The third column displays hodograms taken from the right half of
the simulation domains. The dotted curves in the second column are the
predicted magnitudes and positions of switch-off slow shocks (SSS) and
fast rarefactions (FR) from isotropic MHD theory.}
\label{LO_degrees} \end{figure}

In order to understand how the temperature anisotropy varies with
other parameters, we perform runs with different upstream angles
$\theta_{BN}$. Fig.~\ref{LO_degrees} documents the results of 
Runs ${\bf a}$, ${\bf b}$, ${\bf c}$, ${\bf d}$, ${\bf f}$ and ${\bf k}$. From the
first column, the downstream $\varepsilon$ tends to lower values in
the more oblique cases and the plasma becomes turbulent once
$\varepsilon$ is comparable to or lower than $\sim 0.25$. We can estimate at which
$\theta_{BN}$ the downstream $\varepsilon$ will drop below 0.25 as
follows.  In the cold plasma limit, the temperature anisotropy due to
the Alfv\'enic counter-streaming ions at the symmetry line is (where
$P_{\|}-P_{\bot} \sim n_a m_i C_{Az}^2$ and $|B|\sim B_{x,a}$),
\begin{align}
\varepsilon_{down} \sim 0.25 \sim 1-\frac{B_{z,a}^2}{B_{x,a}^2} &&
\rightarrow \mbox{tan} ^2\theta_{BN,c} \sim 0.75 && \rightarrow
\theta_{BN,c} \sim 40^\circ
\end{align}
This argument qualitatively shows the tendency to develop stronger
firehose-sense temperature anisotropies for higher obliquities. The
difference between $\theta_{BN,c}$ and the observed value of
$60^\circ$ is probably due to the simplified assumptions, such as a cold
streaming plasma.

In the second column of Fig.~\ref{LO_degrees}, the corresponding
magnetic structures are shown. When the obliquity is large enough,
especially when $\varepsilon <0.25$, the downstream magnetic field
rotates into the out-of-plane direction and becomes
turbulent. Combined with the hodograms in the third column, we can
deduce that the dominant downstream rotational waves are all
left-handed (LH, counter-clockwise in our hodogram). When the
wavelength of the primary LH wave is large, as in the
$\theta_{BN}=60^\circ$ case, its front part breaks into finer
right-handed (RH) waves with scale $\sim 6d_i$. In the $75^\circ$
case, the scale of the primary LH wave is already as small as $6 d_i$,
and so it is more stable than the $60^\circ$ case, albeit still turbulent. (A
$75^\circ$ case with a wider initial current layer is discussed in the
next section. It exhibits wave-generation phenomena similar to the
$60^\circ$ case). In the $83^\circ$ case, we observe RH small-scale
waves in front of the downstream primary LH wave.

For comparison, the dotted curves in the second column of
Fig.~\ref{LO_degrees} are the predicted $B_z$ structure from MHD
theory \cite{lin93a}. The overall predictions agree well for the
oblique cases (see, for instance, the upstream $B_z$ of the slow
shocks in the $60^\circ$ and $75^\circ$ cases), although the reflected
weak fast rarefactions from our boundary have caused a discrepancy in
the slow shock upstream $B_z$ for the $83^\circ$ case. In less oblique
cases, the intermediate and fast characteristic speeds approach one
another just upstream of the switch-off slow shock according to MHD
theory. Therefore there is no clear separation between the slow shocks
upstream and the fast rarefactions, as can be seen in the
simulations. We treat the place where the inflow speed $V_x$ (not
shown) starts to decrease as the upstream of the slow shocks,
which corresponds to the beginning of the LH rotational wavetrains in
the $30^\circ$ and $45^\circ$ cases. Their upstream will hence
correspond to $B_z\sim 0.15$ and $B_z\sim 0.2$ respectively. Therefore
the stable small amplitude rotational waves in the
$\theta_{BN}=30^\circ$ and $45^\circ$ cases are more similar to the
conventional dispersive stationary downstream wavetrains, which immediately follow
the slow shock upstream. (We note that the model in Lin and Lee
\cite{lin93a} approximates the rarefactions by replacing the energy
jump condition in the Rankine-Hugoniot relations by
$[P\rho^{-\gamma}]=0$, arguing that the entropy across a weak
rarefaction does not change. This is only valid for weak rarefaction
waves, but the overall tendency as the propagation angle becomes more
parallel should be in the correct sense).

\begin{figure} 
  \includegraphics[width=13cm]{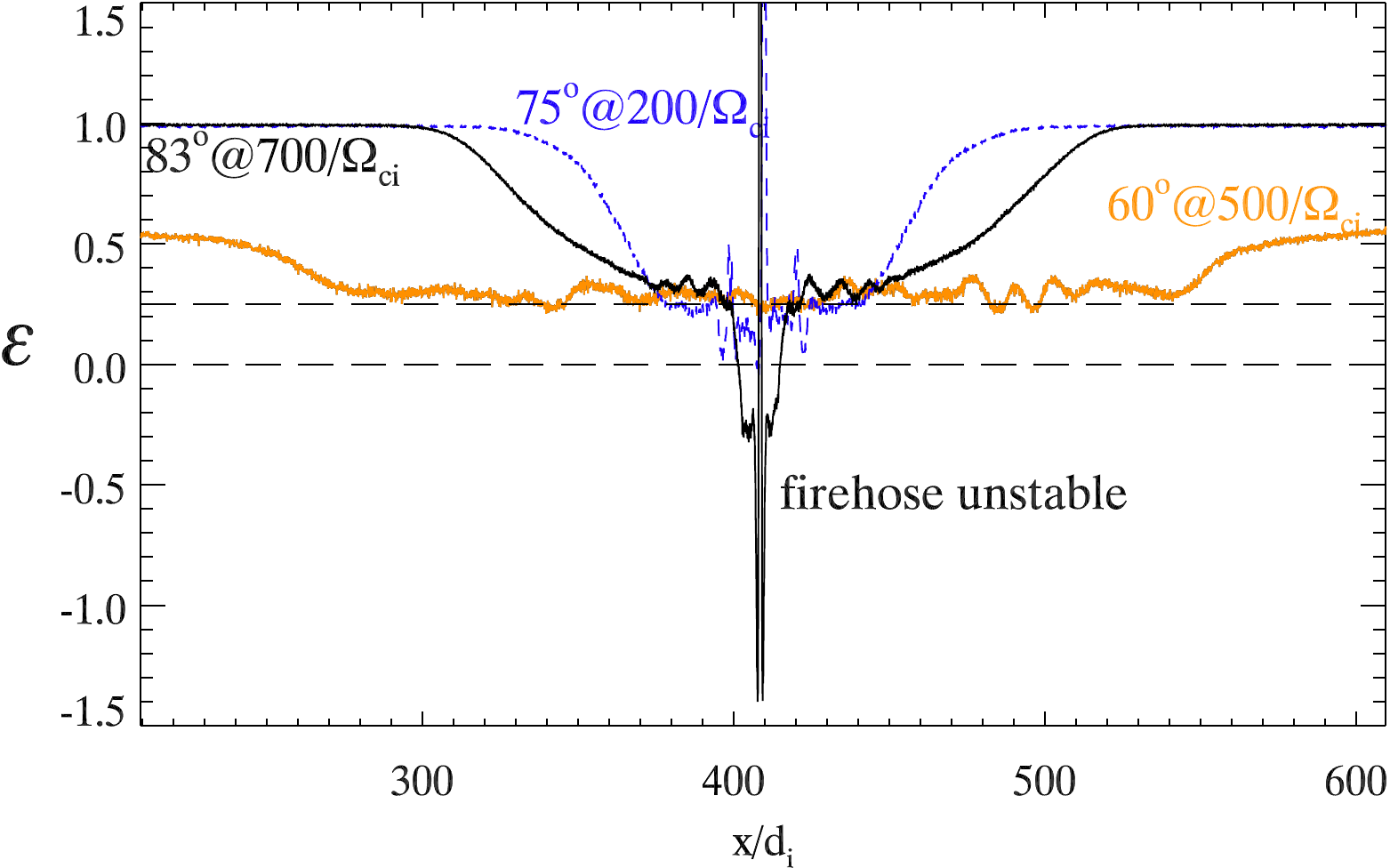} \caption{ The
  $\varepsilon$ distributions of runs $\theta_{BN}=60^\circ$ (Run ${\bf d}$) at
  $500/\Omega_{ci}$, $75^\circ$ (Run ${\bf f}$) at $200/\Omega_{ci}$, $83^\circ$ (Run ${\bf k}$) at
  $700/\Omega_{ci}$. (The $83^\circ$ case is shifted to the right by
  $204.8 d_i$) } \label{LO_epsilons} \end{figure}

An interesting feature in the oblique cases is the coincidence between
the start of the primary LH magnetic rotation and the location where
the anisotropy parameter $\varepsilon \sim 0.25$. The anisotropy
parameter seems to be locked to this critical value for long periods
of time, as is shown in Fig.~\ref{LO_evolve}. Moreover, this is also the
location where turbulence develops. Given these
coincidences the following questions naturally arise. If the
rotational wave is really a normal dispersive wavetrain, why does it
appear in the middle of the MHD predicted switch-off slow shocks? Why
do the dispersive waves not start directly from the slow shock
upstream?  Is it not more similar to a new transition at $B_z\sim 0.3$
for $\theta_{BN} = 60^\circ$, or $B_z\sim 0.5$ for
$\theta_{BN}=75^\circ$ and $83^\circ$?  What is special about
$\varepsilon = 0.25$? The importance of $\varepsilon \sim 0.25$ is shown for different obliquities in Fig. \ref{LO_epsilons}. 
Are there other instabilities associated with
this anisotropy value? Or is it due to the nonlinear structure of a
system with a large temperature anisotropy that cannot be explained by
ideal MHD?

We describe a possible theoretical explanation for why $\varepsilon
\sim 0.25$ in a follow-up paper \cite{yhliu11b}. In short, $\varepsilon=0.25$ represents a
transition where the slow and intermediate mode speeds become
degenerate. Unlike the conventional picture, the downstream rotational
waves can not then be explained by slow dispersive waves, but instead
take the form of rotational intermediate modes.
The coplanar decreasing part of the magnetic field together with the non-coplanar rotational part will later be identified as a single nonlinear wave, called a compound SS/RD wave \cite{yhliu11b}.

\section{The downstream turbulent waves and particle scattering}\

\begin{figure} 
\includegraphics[width=12cm]{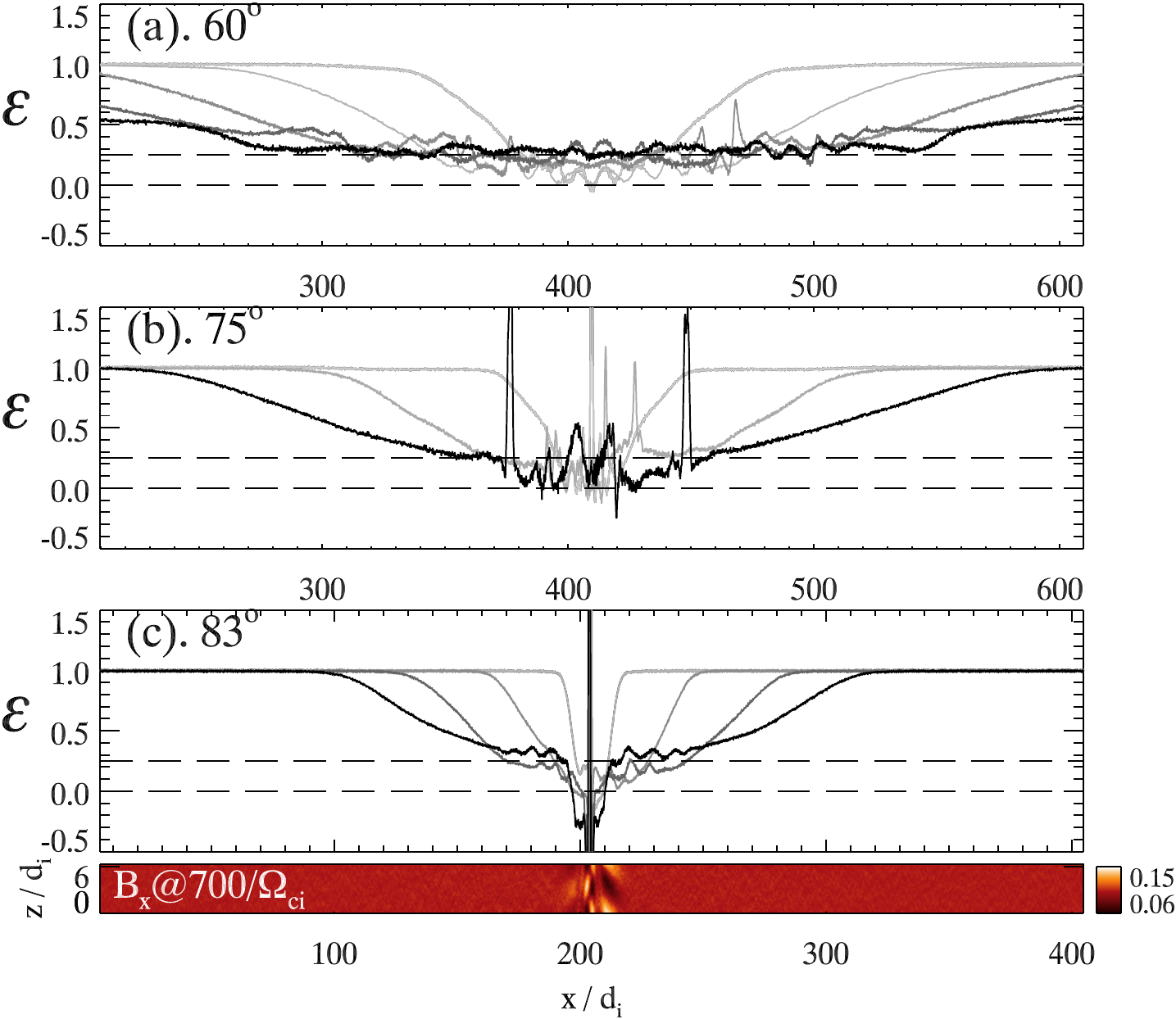} 
\caption{Evolution of $\varepsilon$ for the case with
$\theta_{BN}=60^\circ$(Run ${\bf d}$) for equally spaced times between $100-500/\Omega_{ci}$ from lighter
grey to darker grey in (a), the $\theta_{BN}=75^\circ$ case (Run ${\bf f}$) for time
$100-500/\Omega_{ci}$ in (b), and the $\theta_{BN}=83^\circ$ case (Run ${\bf k}$) for time
$100-700/\Omega_{ci}$ in (c). The bottom is a plot of $B_x$ for the
$\theta_{BN}=83^\circ$ case at time $700/\Omega_{ci}$ showing the 2-D turbulence that develops. }
\label{LO_evolve} \end{figure}

The region downstream of oblique ({$\theta_{BN} \geq 60^\circ$}) slow
shocks is turbulent. High wavenumber waves are continually
excited whenever the $B_z$ component begins to rotate into the
out-of-plane direction, as can be seen for the $60^\circ$, $75^\circ$
and $83^\circ$ cases in Fig.~\ref{LO_degrees}, and more clearly in the
evolution of the $B$ field for the case $75^\circ,w_i=10d_i$ (Run ${\bf g}$) in
Fig.~\ref{LO_cascade}. The downstream LH rotational parent waves break
into $\lambda_x \sim 6d_i$-scale waves. The large oscillation in
$\varepsilon$ at later times is due to the small magnetic field
magnitude near the symmetry line, where both $B_z$ and $B_y$ vanish.
The particle scattering associated with these small-scale waves plays
a role in counter-balancing the decrease in $\varepsilon$ due to the
streaming ions and keeping the temperature anisotropy around the value
0.25, as is seen in the time evolution of the oblique cases in
Fig.~\ref{LO_evolve}.

In order to understand the downstream turbulent $d_i$-scale waves, we
tried to systematically pin down the possible driver and energy source
via numerical experiments. We separately checked potential energy sources 
in the shock simulations, including temperature anisotropy,
counterstreaming beams, rotational parent waves (such as the larger LH
wave in Fig.~\ref{LO_cascade}) to determine which factors are responsible
for generating the $d_i$-scale waves.

We carried out spatially homogeneous simulations with an initial wave structure of the following form:
\begin{equation}
B_z=B_{\text{cir}}\cos(2\pi x/\lambda_p)+B_{z,\text{oblique}}; \qquad B_y=P \times
B_{\text{cir}}\sin(2 \pi x/\lambda_p)
\end{equation} 
When $B_{\text{cir}} \neq 0$, there is a rotational field with $P=+1$
for LH, $-1$ for RH, and $0$ for planar polarizations. The constant
$B_{z,\text{oblique}}$ controls the obliquity of this circularly
polarized wave and provides a spatial modulation in the magnitude of
the total magnetic field. The general expression for the initial ion
distribution is
\begin{equation}
 f_i \propto \mbox{exp}\left(-\frac{m_i (v_\|-u)^2}{2T_{i\|}}-\frac{m_i v_\bot^2}{T_{i\bot}}\right)+\mbox{exp}\left(-\frac{m_i (v_\|+u)^2}{2T_{i\|}}-\frac{m_i v_\bot^2}{T_{i\bot}}\right)
\end{equation}
which has bi-Maxwellian counterstreaming beams for $u \neq 0$. Note
that $T_{i\|,\text{eff}}=T_{i\|}+m_i u^2$, so both $u\neq0$ and
$T_{i\|} \neq T_{i\bot}$ can contribute to the temperature anisotropy
$\varepsilon$. The initial plasma density varies so as to ensure a
constant value of $P_\bot+B^2/(2\mu_0)$. Since the small-scale waves
that interest us do not induce variation in the $z$-direction, they
are intrinsically 1-D waves along the $x$-direction. The common
parameters used here are a domain size of $1.6
\times 51.2$, with grid $64 \times 2048$, $\lambda_p=51.2 d_i$, uniform
$T_{i,e\bot}=T_{e\|}=0.15$, $B_x=0.25$ and $n_{i,e}=1.5$ at $x=0$. These
parameters are meant to represent those observed in the downstream of
the $75^\circ$ case.\newline

\begin{table}
\caption{}
\begin{center}
\begin{tabular}{ccccccccc}
\tableline\tableline
Run & $B_{cir}$ & Polarization(P) & $B_{z,oblique}$ & $T_{i\|}$ & Beams (u) &$\rightarrow$\footnotemark[1] $T_{i\|,eff}$ &$\rightarrow \varepsilon$& $\rightarrow$ 1-D $d_i$-scale waves ($t <100/\Omega_{ci}$)\\
\tableline
${\bf 1}$ & 0.25 & +1 & 0.25 & 0.15& 0.5&0.4& -0.2 $\sim$ -7 & $\surd$\\
${\bf 2}$ & 0.25 & +1 & 0.25 & 0.4& 0 &0.4& -0.2 $\sim$ -7 & $\surd$\\
${\bf 3}$ & 0.25 & +1 & 0 & 0.15& 0.5& 0.4& -2 & x\\
${\bf 4}$ & 0.25 & +1 & 0 & 0.4 & 0 &0.4& -2 & x\\
${\bf 5}$ & 0.25 & +1 & 0.25 & 0.15 & 0&0.15 & 1 & $\surd$\\
${\bf 6}$  & 0.25 & 0 & 0.25 & 0.15 & 0 &0.15 & 1 & x\\
\tableline
\end{tabular}
\footnotetext[1]{``$\rightarrow$'' means ``resulting in''.}
\end{center}
\end{table}

\begin{figure} 
   \includegraphics[width=12cm]{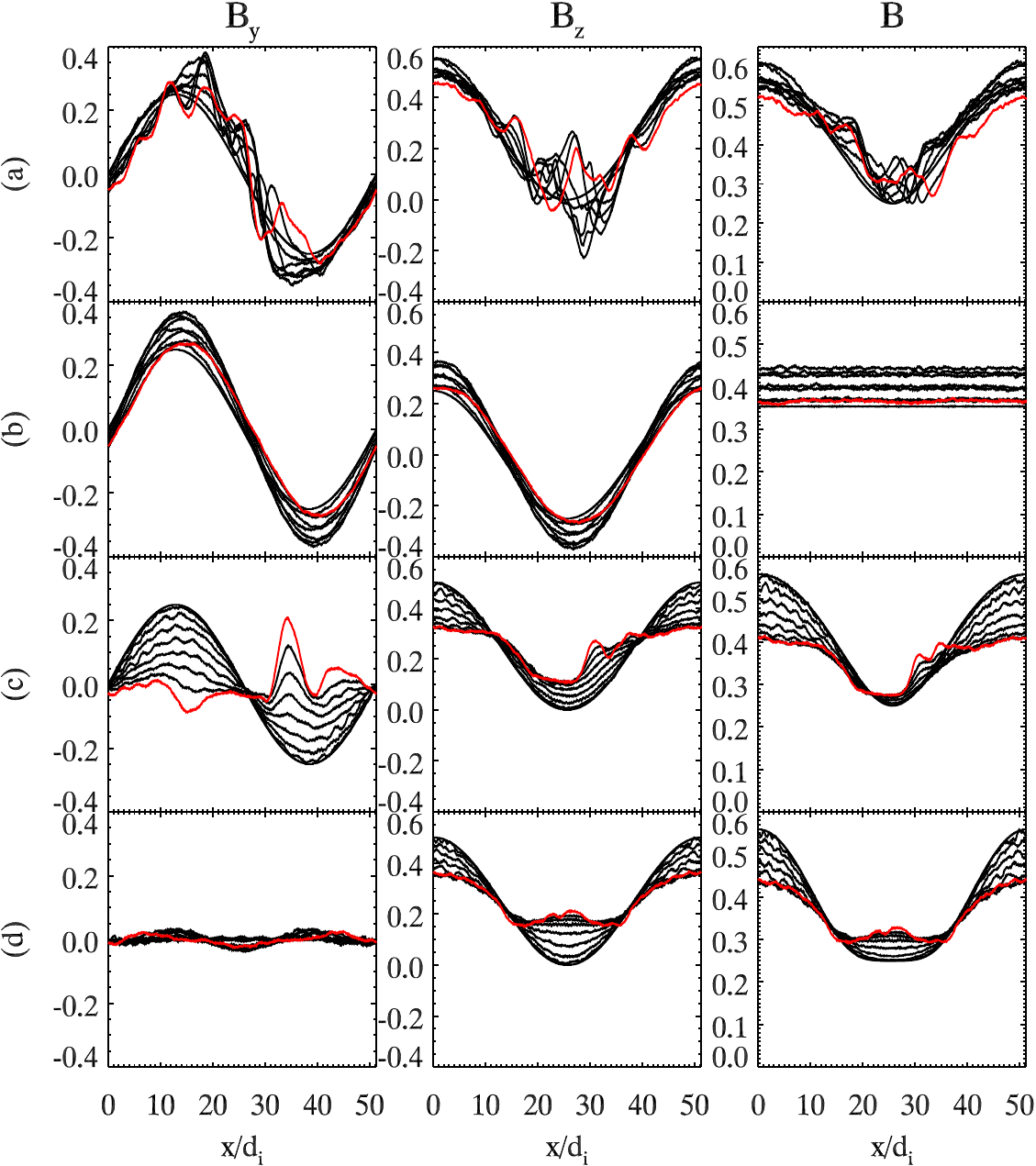} 
\caption{The evolution of $B_y$, $B_z$ and $B$ for equally spaced times between $0 -
100/\Omega_{ci}$.  The red curve indicates the time
$100/\Omega_{ci}$. Panel (a): Run ${\bf 1}$ with both initial
streaming ions and modulated rotational parent wave. Panel (b): The
same as panel (a) without the initial spatial modulation (Run ${\bf
3}$). Panel (c): The same as panel (a) without initial beams (Run
${\bf 5}$). Panel (d): The same as panel (c) without initial
polarization (Run ${\bf 6}$). }
\label{LO_MI}
\end{figure}

Run ${\bf 1}$ of Table II is a representative example of the
downstream structure seen in the Riemann simulations. The initial
obliquely propagating LH polarized waves with Alfv\'enic
counter-streaming ions along the local magnetic field break into $\sim
6d_i$ small-scale waves, as shown in Fig.~\ref{LO_MI}(a). In Run ${\bf
2}$, we replace the counter-streaming ions with a bi-Maxwellian plasma
with the same effective parallel temperature, therefore the same
temperature anisotropy $\varepsilon$, and find similar wave-generation
phenomena. In Runs ${\bf 3}$ and ${\bf 4}$, we remove the obliquity
($B_{z,\text{oblique}}$) from Runs ${\bf 1}$ and ${\bf 2}$
respectively. No $d_i$-scale waves are excited within a time of
$100/\Omega_{ci}$ in Run ${\bf 3}$, as is shown in
Fig.~\ref{LO_MI}(b). This indicates the importance of spatial
modulation for the development of the turbulence. In Run ${\bf 5}$, we
replace the beams or anisotropic plasma of Runs ${\bf 1}$ and ${\bf
2}$, respectively, with a Maxwellian isotropic plasma. Waves with
scale $\sim 6d_i$ are excited and are shown in
Fig.~\ref{LO_MI}(c). However, if we further remove the initial
out-of-plane magnetic field of the parent wave (Run ${\bf 6}$), no
small-scale waves appear; see Fig.~\ref{LO_MI}(d). This indicates that
a necessary condition for producing these small-scale waves is the
existence of circularly polarized parent waves.  From this suite of
runs we conclude that the presence of a spatially modulated rotational
wave is the major driver of $d_i$-scale coherent waves seen downstream
of the shock (see Fig.~\ref{LO_cascade}(b)). Although not shown, we
see similar behvior for RH ($P=-1$) parent waves.

A non-modulated, constant-amplitude, circularly polarized Alfv\'en
wave is a known stationary solution of the MHD equations. Although it
is subject to a long-wavelength modulational instability under some
conditions \cite{mjolhus76a}, simulations show that it is also stable
in a collisionless plasma. Spangler \cite{spangler92a} explored a
similar wave-generation phenomena, as well as the formation of
solitary waves with a system with initial conditions similar to our
Run ${\bf 5}$.  He cast his system in terms of the
Derivative-Nonlinear-Schr\"odinger-Equations, which are simplified MHD
equations that possess the MHD nonlinearity and dispersion terms. This
suggests the importance of both the nonlinearity and the Hall
dispersion term for these $d_i$-scale waves. We also note that the
time of onset of these waves is proportional to $\lambda_p$ and
inversely proportional to the amplitude $B_{\text{cir}}$, which are
closely related to the steepening time scale of finite amplitude
waves, $T_s \sim (\tau_{wave}/2\pi)(B/\Delta B)$, with $\tau_{wave}
\propto \lambda_p$ (the gradient scale of the finite amplitude wave)
\cite{barnes74a}. Although a firehose-sense temperature anisotropy
($\varepsilon <0$) would weaken the nonlinearity, the spatial
variation of $\varepsilon$ enhances the nonlinearity, as will be shown
in the follow-up paper \cite{yhliu11b}.

\begin{figure} \includegraphics[width=10cm]{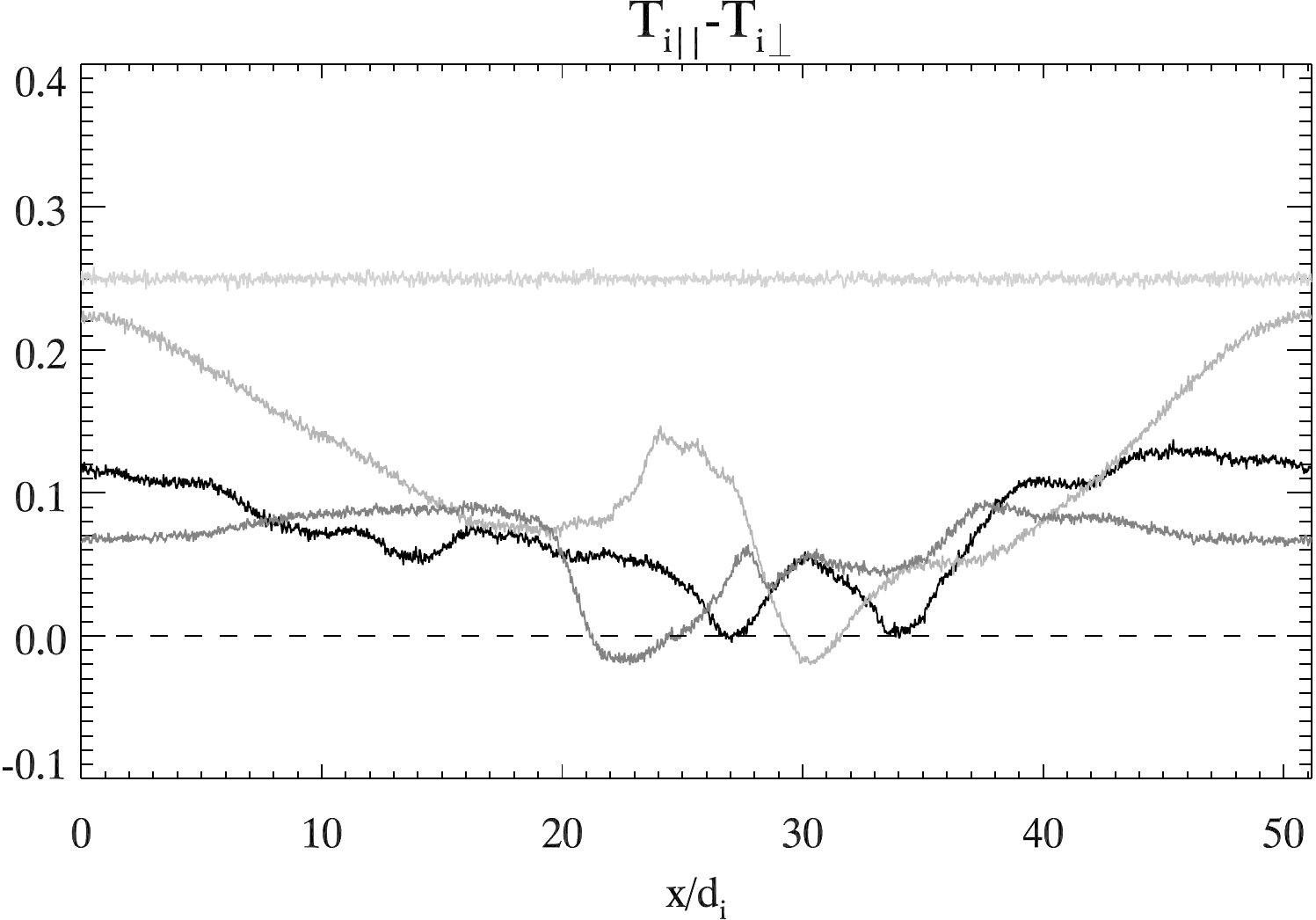} 
\caption{The evolution of $T_{i\|}-T_{i\bot}$ for equally spaced times between $0 -
100/\Omega_{ci}$(from lighter grey to darker grey) of Run ${\bf 1} $
(Fig.~\ref{LO_MI}(a)). The temperature anisotropy of the ions is
reduced, which indicates particle scattering is taking place.}
\label{LO_scatter} \end{figure}

Runs ${\bf 3}$ and ${\bf 4}$ suggest that temperature anisotropy and
Alfv\'enic beams are not, by themselves, sufficient for generating the
small-scale waves seen in the downstream region. The time scale of the
wave-generation is faster than beam/anisotropy driven waves from the
electromagnetic ion-ion cyclotron instability (EMIIC), such as
Alfv\'en/ion cyclotron (AIC) waves \cite{winske92a}, or
kinetic-Alfv\'en waves (KAW)\cite{yin07a}. As is the case for the
growth rate of the firehose instability $\gamma^2 \sim (-\varepsilon
k^2C_A^2)/(1+C_A^2/c^2) $ \cite{davidson68a}, the growth rate is small
when $C_A$ (based on the ambient B field) is small. However, these
beam/anisotropy-driven waves are still potential players in the
downstream turbulence at late time. By comparing Fig.~\ref{LO_MI}(a)
to (c), we see that the interplay of the steepening process and the
free-streaming beams makes the wave more turbulent. In
Fig.~\ref{LO_scatter}, the ion temperature anisotropy
$T_{i\|}-T_{i\bot}$ of Run ${\bf 1}$ (which corresponds to
Fig.~\ref{LO_MI}(a)) decreases. This confirms the ability of these
smaller-scale dispersive waves to scatter ions.

The polarization of a linear wave can be determined by the phase
between the tangential magnetic field variations $\delta B_z$ and
$\delta B_y$. In the fluid model $i \delta B_z/\delta B_y=
(C_I^2-(\omega/k)^2)/(\omega d_i C_A \mbox{cos}(\theta_{BN}))$
\cite{walthour94a}. A wave has a LH polarization when
$(\omega/k)^2 > C_I^2$, and RH otherwise. As observed in our oblique
shock simulations (discussed in Sec.~\ref{ta_section}), the
small-scale waves at the upstream of the rotational front are mostly
RH, while the downstream waves are mostly LH. This suggests that the
primary rotational front propagates at the local intermediate speed
(as also measured in the simulations).  We do not address which modes
are responsible for these fine-scale waves here, since the
polarization of the linear mode in kinetic theory is very complicated.
Both temperature anisotropy and high plasma $\beta$ play roles in
changing the linear wave properties
\cite{krauss-varban94a}.  With a larger $l_z$  and very oblique
propagation angle, as in the $83^\circ$ case, 2-D turbulence with
$\lambda_z \sim 6d_i$ is excited around the firehose-unstable
region. Its signature is most clear in $B_x$, as can be seen in the
bottom plot of Fig.~\ref{LO_evolve}. A similar mechanism is postulated
to excite 2-D turbulence in the firehose-unstable region seen in the
reconnection simulation of Fig.~\ref{LO_reconn}(c) (and, perhaps, the
2-D turbulence previously reported \cite{scholer98a}). This 2-D turbulence could be driven 
by temperature anisotropy (firehose-like) or by the sharp front of the primary rotational wave.

\section{summary and discussion}\

We have studied the temperature anisotropy distribution across slow
shocks with different obliquities in PIC simulations.  An abnormal
transition and an anisotropy $\varepsilon=0.25$ locking phenomena
downstream of the MHD predicted switch-off slow shocks is
documented. The Alfv\'enic counter-streaming ions serve as the driver
for decreasing $\varepsilon$ (increasing the firehose-sense
temperature anisotropy) in the downstream region, while downstream
$d_i$-scale turbulent waves scatter particles and raise
$\varepsilon$. This dynamical balance makes the downstream
$\varepsilon$ plateau at a value of $\varepsilon=0.25$ and not the
marginal firehose criterion $\varepsilon=0$. The theoretical
significance of $\varepsilon= 0.25$ will be addressed in another work
\cite{yhliu11b}. By means of PIC numerical experiments we show that
the turbulent $d_i$-scale waves are radiated from a spatially
modulated rotational parent wave .

\begin{figure}
\includegraphics[width=12cm]{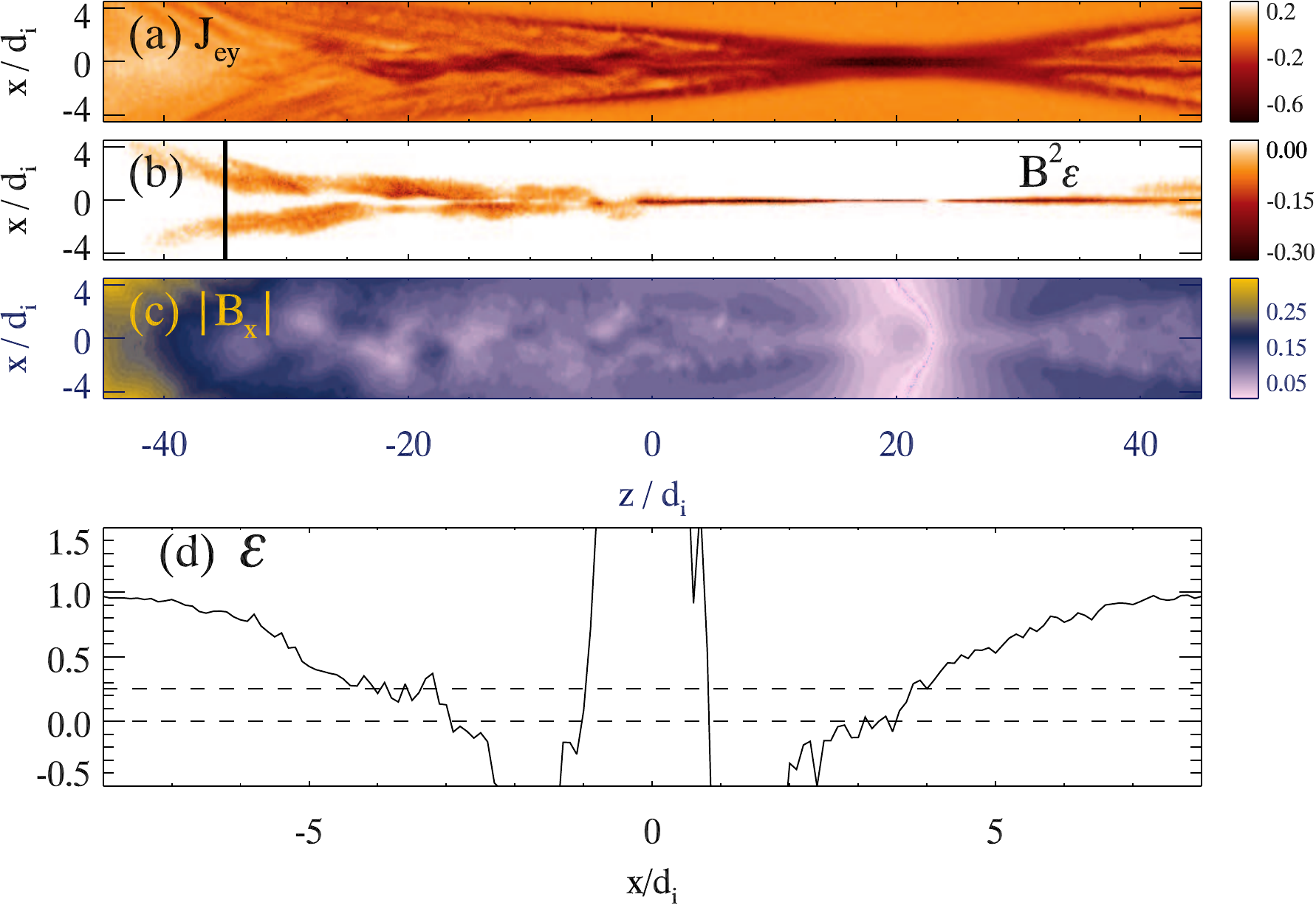} 
\caption{ The exhaust from steady reconnection in a PIC simulation. Panel (a): The
out-of-plane electron current $J_{ey}$; Panel (b): $B^2\varepsilon$,
where positive values have been set to 0. The colored region is
firehose unstable; Panel (c): The magnitude of $B_x$ showing the
turbulence associated with the firehose instability; Panel (d): A cut
of $\varepsilon$ at $x/d_i=-35$ (the vertical line in (b)).  The
horizontal lines demark $\varepsilon = 0.25$ and $\varepsilon = 0$.}
\label{LO_reconn} \end{figure}

The Riemann problem for the $\theta_{BN}=83^\circ$ case is closely
related to that of reconnection exhausts with normalized reconnection
rates of $0.1$. The very center is a firehose-unstable region where a
$B_x$ variation is observed, as in the reconnection
simulations. Although we cannot confidently identify the $0.25$
plateau in present PIC reconnection simulations (since the spatial
extension in the normal direction ($\hat{{\bf e}}_x$), is $\sim 10
d_i$; see Fig.~\ref{LO_reconn}(d)), we expect to see the signature of
a $\varepsilon=0.25$ plateau outside the firehose unstable region in
very large kinetic anti-parallel reconnection simulations and in-situ
satellite observations of anti-parallel magnetic reconnection
outflows.

\begin{figure} 
    \includegraphics[width=16cm]{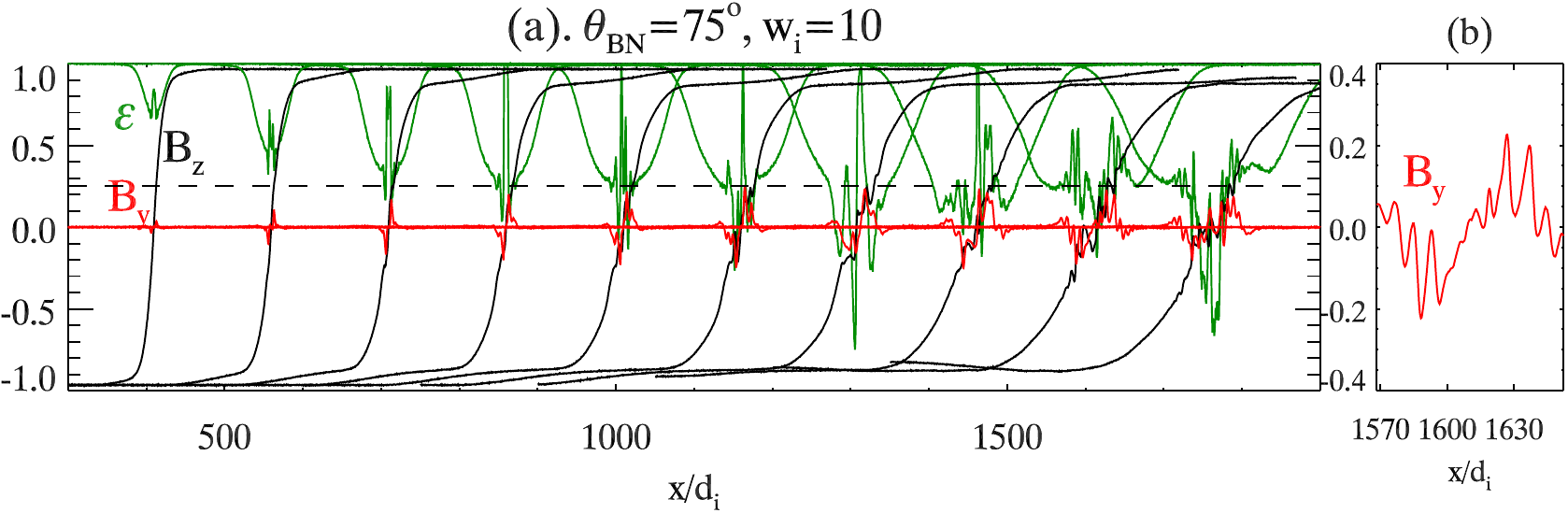} 
\caption{Panel (a): The evolution of $\varepsilon$, $B_z$ and $B_y$ for equally spaced
times between $50 - 500/\Omega_{ci}$ (from left to right) in the $\theta_{BN}=75^\circ$,
$w_i=10d_i$ case (Run ${\bf g}$). The downstream larger-scale rotational wave breaks
into waves of wavelength $\sim 6d_i$. Panel (b): A blowup of the
downstream $B_y$ at time $450/\Omega_{ci}$. }  \label{LO_cascade}
\end{figure}

Compared to slow shocks, fast shocks have been intensively studied
(see, for instance, a review article \cite{treumann09a}). The formation of shocklet and
short large-amplitude magnetic structures (SLAMs) in front of earth's
quasi-parallel bow shock has been observed \cite{lucek02a} and
simulated \cite{omidi90a,scholer93a}. In our slow shock simulations,
the development of downstream rotational waves produces SLAMs-like
structures as in Fig.~\ref{LO_cascade}(b). Since the transition of the
magnetic field across the front in a fast shock is opposite to that in
a slow shock, if there are SLAMs associated with a slow shock, they
are expected to exist in the downstream region,  
especially since the downstream of a switch-off slow shock is locally quasi-parallel.

Finally, even though we have not seen super-thermal particles other than
Alfv\'enic streaming ions in our shock simulations, it is still an
open question whether particles could be accelerated by slow shock
associated reconnection exhausts via mechanisms recognized in
fast shocks such as the 1st-order Fermi mechanism \cite{bell78a},
diffusive shock acceleration (DSA) \cite{lee82a,fisk06a}, and those possible 
injection mechanisms for DSA such as shock-drift
\cite{decker88a}, or shock surfing \cite{sagdeev66a}.

\begin{thebibliography}{48}
\expandafter\ifx\csname natexlab\endcsname\relax\def\natexlab#1{#1}\fi
\expandafter\ifx\csname bibnamefont\endcsname\relax
  \def\bibnamefont#1{#1}\fi
\expandafter\ifx\csname bibfnamefont\endcsname\relax
  \def\bibfnamefont#1{#1}\fi
\expandafter\ifx\csname citenamefont\endcsname\relax
  \def\citenamefont#1{#1}\fi
\expandafter\ifx\csname url\endcsname\relax
  \def\url#1{\texttt{#1}}\fi
\expandafter\ifx\csname urlprefix\endcsname\relax\def\urlprefix{URL }\fi
\providecommand{\bibinfo}[2]{#2}
\providecommand{\eprint}[2][]{\url{#2}}

\bibitem[{\citenamefont{Sweet}(1958)}]{sweet58a}
\bibinfo{author}{\bibfnamefont{P.~A.} \bibnamefont{Sweet}}, in
  \emph{\bibinfo{booktitle}{IAU Symp. in Electromagnetic Phenomena in Cosmical
  Physics, ed. B. Lehnet (New York: Cambridge Univ. Press)}}
  (\bibinfo{year}{1958}), p. \bibinfo{pages}{123}.

\bibitem[{\citenamefont{Parker}(1957)}]{parker57a}
\bibinfo{author}{\bibfnamefont{E.~N.} \bibnamefont{Parker}},
  \bibinfo{journal}{J. Geophys. Res.} \textbf{\bibinfo{volume}{62}},
  \bibinfo{pages}{509} (\bibinfo{year}{1957}).

\bibitem[{\citenamefont{Petschek}(1964)}]{petschek64a}
\bibinfo{author}{\bibfnamefont{H.~E.} \bibnamefont{Petschek}}, in
  \emph{\bibinfo{booktitle}{Proc. AAS-NASA Symp. Phys. Solar Flares}}
  (\bibinfo{year}{1964}), vol.~\bibinfo{volume}{50} of
  \emph{\bibinfo{series}{NASA-SP}}, pp. \bibinfo{pages}{425--439}.

\bibitem[{\citenamefont{Tsuneta}(1996)}]{tsuneta96a}
\bibinfo{author}{\bibfnamefont{S.}~\bibnamefont{Tsuneta}},
  \bibinfo{journal}{Astrophys. J.} \textbf{\bibinfo{volume}{456}},
  \bibinfo{pages}{840} (\bibinfo{year}{1996}).

\bibitem[{\citenamefont{Longcope et~al.}(2009)\citenamefont{Longcope, Guidoni,
  and Linton}}]{longcope09a}
\bibinfo{author}{\bibfnamefont{D.~W.} \bibnamefont{Longcope}},
  \bibinfo{author}{\bibfnamefont{S.~E.} \bibnamefont{Guidoni}},
  \bibnamefont{and} \bibinfo{author}{\bibfnamefont{M.~G.}
  \bibnamefont{Linton}}, \bibinfo{journal}{Astrophys. J.}
  \textbf{\bibinfo{volume}{690}}, \bibinfo{pages}{L18} (\bibinfo{year}{2009}).

\bibitem[{\citenamefont{Feldman et~al.}(1984)\citenamefont{Feldman, Schwartz,
  Bame, Baker, Gosling, E.~W.~Hones, McComas, Slavin, Smith, and
  Zwickl}}]{feldman84a}
\bibinfo{author}{\bibfnamefont{W.~C.} \bibnamefont{Feldman}},
  \bibinfo{author}{\bibfnamefont{S.~J.} \bibnamefont{Schwartz}},
  \bibinfo{author}{\bibfnamefont{S.~J.} \bibnamefont{Bame}},
  \bibinfo{author}{\bibfnamefont{D.~N.} \bibnamefont{Baker}},
  \bibinfo{author}{\bibfnamefont{J.~T.} \bibnamefont{Gosling}},
  \bibinfo{author}{\bibfnamefont{J.}~\bibnamefont{E.~W.~Hones}},
  \bibinfo{author}{\bibfnamefont{D.~J.} \bibnamefont{McComas}},
  \bibinfo{author}{\bibfnamefont{J.~A.} \bibnamefont{Slavin}},
  \bibinfo{author}{\bibfnamefont{E.~J.} \bibnamefont{Smith}}, \bibnamefont{and}
  \bibinfo{author}{\bibfnamefont{R.~D.} \bibnamefont{Zwickl}},
  \bibinfo{journal}{Geophys. Res. Lett.} \textbf{\bibinfo{volume}{11}},
  \bibinfo{pages}{599} (\bibinfo{year}{1984}).

\bibitem[{\citenamefont{Feldman et~al.}(1987)\citenamefont{Feldman, Tokar,
  Birn, E.~W.~Hones, Bame, and Russell}}]{feldman87a}
\bibinfo{author}{\bibfnamefont{W.~C.} \bibnamefont{Feldman}},
  \bibinfo{author}{\bibfnamefont{R.~L.} \bibnamefont{Tokar}},
  \bibinfo{author}{\bibfnamefont{J.}~\bibnamefont{Birn}},
  \bibinfo{author}{\bibfnamefont{J.}~\bibnamefont{E.~W.~Hones}},
  \bibinfo{author}{\bibfnamefont{S.~J.} \bibnamefont{Bame}}, \bibnamefont{and}
  \bibinfo{author}{\bibfnamefont{C.~T.} \bibnamefont{Russell}},
  \bibinfo{journal}{J. Geophys. Res.} \textbf{\bibinfo{volume}{92}},
  \bibinfo{pages}{83} (\bibinfo{year}{1987}).

\bibitem[{\citenamefont{Smith et~al.}(1984)\citenamefont{Smith, Slavin,
  Tsurutani, Feldman, and Bame}}]{smith84a}
\bibinfo{author}{\bibfnamefont{E.~J.} \bibnamefont{Smith}},
  \bibinfo{author}{\bibfnamefont{J.~A.} \bibnamefont{Slavin}},
  \bibinfo{author}{\bibfnamefont{B.~T.} \bibnamefont{Tsurutani}},
  \bibinfo{author}{\bibfnamefont{W.~C.} \bibnamefont{Feldman}},
  \bibnamefont{and} \bibinfo{author}{\bibfnamefont{S.~J.} \bibnamefont{Bame}},
  \bibinfo{journal}{Geophys. Res. Lett.} \textbf{\bibinfo{volume}{11}},
  \bibinfo{pages}{1054} (\bibinfo{year}{1984}).

\bibitem[{\citenamefont{{\O}ieroset et~al.}(2000)\citenamefont{{\O}ieroset,
  Phan, Lin, and Sonnerup}}]{oieroset00a}
\bibinfo{author}{\bibfnamefont{M.}~\bibnamefont{{\O}ieroset}},
  \bibinfo{author}{\bibfnamefont{T.~D.} \bibnamefont{Phan}},
  \bibinfo{author}{\bibfnamefont{R.~P.} \bibnamefont{Lin}}, \bibnamefont{and}
  \bibinfo{author}{\bibfnamefont{B.~U.~{\"O}.} \bibnamefont{Sonnerup}},
  \bibinfo{journal}{J. Geophys. Res.} \textbf{\bibinfo{volume}{105}},
  \bibinfo{pages}{25247} (\bibinfo{year}{2000}).
\bibitem[{\citenamefont{Hoshino et~al.}(2000)\citenamefont{Hoshino, Mukai,
  Shinohara, Saito, and Kokubun}}]{hoshino00a}
\bibinfo{author}{\bibfnamefont{M.}~\bibnamefont{Hoshino}},
  \bibinfo{author}{\bibfnamefont{T.}~\bibnamefont{Mukai}},
  \bibinfo{author}{\bibfnamefont{I.}~\bibnamefont{Shinohara}},
  \bibinfo{author}{\bibfnamefont{Y.}~\bibnamefont{Saito}}, \bibnamefont{and}
  \bibinfo{author}{\bibfnamefont{S.}~\bibnamefont{Kokubun}},
  \bibinfo{journal}{J. Geophys. Res.} \textbf{\bibinfo{volume}{105}},
  \bibinfo{pages}{337} (\bibinfo{year}{2000}).

\bibitem[{\citenamefont{Walthour et~al.}(1994)\citenamefont{Walthour, Gosling,
  Sonnerup, and Russell}}]{walthour94a}
\bibinfo{author}{\bibfnamefont{D.~W.} \bibnamefont{Walthour}},
  \bibinfo{author}{\bibfnamefont{J.~T.} \bibnamefont{Gosling}},
  \bibinfo{author}{\bibfnamefont{B.~U.~{\"O}.} \bibnamefont{Sonnerup}},
  \bibnamefont{and} \bibinfo{author}{\bibfnamefont{C.~T.}
  \bibnamefont{Russell}}, \bibinfo{journal}{J. Geophys. Res.}
  \textbf{\bibinfo{volume}{99}}, \bibinfo{pages}{23705} (\bibinfo{year}{1994}).

\bibitem[{\citenamefont{Swift}(1983)}]{swift83a}
\bibinfo{author}{\bibfnamefont{D.~W.} \bibnamefont{Swift}},
  \bibinfo{journal}{J. Geophys. Res.} \textbf{\bibinfo{volume}{88}},
  \bibinfo{pages}{5685} (\bibinfo{year}{1983}).

\bibitem[{\citenamefont{Winske et~al.}(1985)\citenamefont{Winske, Stover, and
  Gary}}]{winske85a}
\bibinfo{author}{\bibfnamefont{D.}~\bibnamefont{Winske}},
  \bibinfo{author}{\bibfnamefont{E.~K.} \bibnamefont{Stover}},
  \bibnamefont{and} \bibinfo{author}{\bibfnamefont{S.~P.} \bibnamefont{Gary}},
  \bibinfo{journal}{Geophys. Res. Lett.} \textbf{\bibinfo{volume}{12}},
  \bibinfo{pages}{295} (\bibinfo{year}{1985}).

\bibitem[{\citenamefont{Omidi and Winske}(1992)}]{omidi92a}
\bibinfo{author}{\bibfnamefont{N.}~\bibnamefont{Omidi}} \bibnamefont{and}
  \bibinfo{author}{\bibfnamefont{D.}~\bibnamefont{Winske}},
  \bibinfo{journal}{J. Geophys. Res.} \textbf{\bibinfo{volume}{97}},
  \bibinfo{pages}{14801} (\bibinfo{year}{1992}).

\bibitem[{\citenamefont{Lin and Lee}(1993)}]{lin93a}
\bibinfo{author}{\bibfnamefont{Y.}~\bibnamefont{Lin}} \bibnamefont{and}
  \bibinfo{author}{\bibfnamefont{L.~C.} \bibnamefont{Lee}},
  \bibinfo{journal}{Space Science Reviews} \textbf{\bibinfo{volume}{65}},
  \bibinfo{pages}{1} (\bibinfo{year}{1993}).
\bibitem[{\citenamefont{Yin et~al.}(2005)\citenamefont{Yin, Winske, Daughton,
  and Coroniti}}]{yin05a}
\bibinfo{author}{\bibfnamefont{L.}~\bibnamefont{Yin}},
  \bibinfo{author}{\bibfnamefont{D.}~\bibnamefont{Winske}},
  \bibinfo{author}{\bibfnamefont{W.}~\bibnamefont{Daughton}}, \bibnamefont{and}
  \bibinfo{author}{\bibfnamefont{F.~V.} \bibnamefont{Coroniti}},
  \bibinfo{journal}{J. Geophys. Res.} \textbf{\bibinfo{volume}{110}}
  (\bibinfo{year}{2005}).

\bibitem[{\citenamefont{Yin et~al.}(2007{\natexlab{a}})\citenamefont{Yin,
  Winske, and Daughton}}]{yin07b}
\bibinfo{author}{\bibfnamefont{L.}~\bibnamefont{Yin}},
  \bibinfo{author}{\bibfnamefont{D.}~\bibnamefont{Winske}}, \bibnamefont{and}
  \bibinfo{author}{\bibfnamefont{W.}~\bibnamefont{Daughton}},
  \bibinfo{journal}{Phys. Plasmas} \textbf{\bibinfo{volume}{14}},
  \bibinfo{eid}{062105} (\bibinfo{year}{2007}{\natexlab{a}}).

\bibitem[{\citenamefont{Drake et~al.}(2009)\citenamefont{Drake, Swisdak, Phan,
  Cassak, Shay, Lepri, Lin, Quataert, and Zurbuchen}}]{drake09a}
\bibinfo{author}{\bibfnamefont{J.~F.} \bibnamefont{Drake}},
  \bibinfo{author}{\bibfnamefont{M.}~\bibnamefont{Swisdak}},
  \bibinfo{author}{\bibfnamefont{T.~D.} \bibnamefont{Phan}},
  \bibinfo{author}{\bibfnamefont{P.~A.} \bibnamefont{Cassak}},
  \bibinfo{author}{\bibfnamefont{M.~A.} \bibnamefont{Shay}},
  \bibinfo{author}{\bibfnamefont{S.~T.} \bibnamefont{Lepri}},
  \bibinfo{author}{\bibfnamefont{R.~P.} \bibnamefont{Lin}},
  \bibinfo{author}{\bibfnamefont{E.}~\bibnamefont{Quataert}}, \bibnamefont{and}
  \bibinfo{author}{\bibfnamefont{T.~H.} \bibnamefont{Zurbuchen}},
  \bibinfo{journal}{J. Geophys. Res.} \textbf{\bibinfo{volume}{114}},
  \bibinfo{pages}{05111} (\bibinfo{year}{2009}).

\bibitem[{\citenamefont{Bale et~al.}(2009)\citenamefont{Bale, Kasper, Howes,
  Quataert, Salem, and Sundkvist}}]{bale09a}
\bibinfo{author}{\bibfnamefont{S.}~\bibnamefont{Bale}},
  \bibinfo{author}{\bibfnamefont{J.}~\bibnamefont{Kasper}},
  \bibinfo{author}{\bibfnamefont{G.~G.} \bibnamefont{Howes}},
  \bibinfo{author}{\bibfnamefont{E.}~\bibnamefont{Quataert}},
  \bibinfo{author}{\bibfnamefont{C.}~\bibnamefont{Salem}}, \bibnamefont{and}
  \bibinfo{author}{\bibfnamefont{D.}~\bibnamefont{Sundkvist}},
  \bibinfo{journal}{Phys. Rev. Lett.} \textbf{\bibinfo{volume}{103}},
  \bibinfo{pages}{211101} (\bibinfo{year}{2009}).
\bibitem[{\citenamefont{Lottermoser et~al.}(1998)\citenamefont{Lottermoser,
  Scholer, and Matthews}}]{lottermoser98a}
\bibinfo{author}{\bibfnamefont{R.~F.} \bibnamefont{Lottermoser}},
  \bibinfo{author}{\bibfnamefont{M.}~\bibnamefont{Scholer}}, \bibnamefont{and}
  \bibinfo{author}{\bibfnamefont{A.~P.} \bibnamefont{Matthews}},
  \bibinfo{journal}{J. Geophys. Res.} \textbf{\bibinfo{volume}{103}},
  \bibinfo{pages}{4547} (\bibinfo{year}{1998}).

\bibitem[{\citenamefont{Nakamura et~al.}(1998)\citenamefont{Nakamura, Fujimoto,
  and Maezawa}}]{nakamura98a}
\bibinfo{author}{\bibfnamefont{M.~S.} \bibnamefont{Nakamura}},
  \bibinfo{author}{\bibfnamefont{M.}~\bibnamefont{Fujimoto}}, \bibnamefont{and}
  \bibinfo{author}{\bibfnamefont{K.}~\bibnamefont{Maezawa}},
  \bibinfo{journal}{J. Geophys. Res.} \textbf{\bibinfo{volume}{103}},
  \bibinfo{pages}{4531} (\bibinfo{year}{1998}).

\bibitem[{\citenamefont{Lin and Swift}(1996)}]{lin96a}
\bibinfo{author}{\bibfnamefont{Y.}~\bibnamefont{Lin}} \bibnamefont{and}
  \bibinfo{author}{\bibfnamefont{D.~W.} \bibnamefont{Swift}},
  \bibinfo{journal}{J. Geophys. Res.} \textbf{\bibinfo{volume}{101}},
  \bibinfo{pages}{19859} (\bibinfo{year}{1996}).

\bibitem[{\citenamefont{Scholer and Lottermoser}(1998)}]{scholer98a}
\bibinfo{author}{\bibfnamefont{M.}~\bibnamefont{Scholer}} \bibnamefont{and}
  \bibinfo{author}{\bibfnamefont{R.~F.} \bibnamefont{Lottermoser}},
  \bibinfo{journal}{Geophys. Res. Lett.} \textbf{\bibinfo{volume}{25}},
  \bibinfo{pages}{3281} (\bibinfo{year}{1998}).

\bibitem[{\citenamefont{Cremer and Scholer}(1999)}]{cremer99a}
\bibinfo{author}{\bibfnamefont{M.}~\bibnamefont{Cremer}} \bibnamefont{and}
  \bibinfo{author}{\bibfnamefont{M.}~\bibnamefont{Scholer}},
  \bibinfo{journal}{Geophys. Res. Lett.} \textbf{\bibinfo{volume}{26}},
  \bibinfo{pages}{2709} (\bibinfo{year}{1999}).

\bibitem[{\citenamefont{Cremer and Scholer}(2000)}]{cremer00a}
\bibinfo{author}{\bibfnamefont{M.}~\bibnamefont{Cremer}} \bibnamefont{and}
  \bibinfo{author}{\bibfnamefont{M.}~\bibnamefont{Scholer}},
  \bibinfo{journal}{J. Geophys. Res.} \textbf{\bibinfo{volume}{105}},
  \bibinfo{pages}{27621} (\bibinfo{year}{2000}).

\bibitem[{\citenamefont{Zeiler et~al.}(2002)\citenamefont{Zeiler, Biskamp,
  Drake, Rogers, Shay, and Scholer}}]{zeiler02a}
\bibinfo{author}{\bibfnamefont{A.}~\bibnamefont{Zeiler}},
  \bibinfo{author}{\bibfnamefont{D.}~\bibnamefont{Biskamp}},
  \bibinfo{author}{\bibfnamefont{J.~F.} \bibnamefont{Drake}},
  \bibinfo{author}{\bibfnamefont{B.~N.} \bibnamefont{Rogers}},
  \bibinfo{author}{\bibfnamefont{M.~A.} \bibnamefont{Shay}}, \bibnamefont{and}
  \bibinfo{author}{\bibfnamefont{M.}~\bibnamefont{Scholer}},
  \bibinfo{journal}{J. Geophys. Res.} \textbf{\bibinfo{volume}{107}},
  \bibinfo{pages}{1230} (\bibinfo{year}{2002}).

\bibitem[{\citenamefont{Coroniti}(1971)}]{coroniti71a}
\bibinfo{author}{\bibfnamefont{F.~V.} \bibnamefont{Coroniti}},
  \bibinfo{journal}{Nucl. Fusion} \textbf{\bibinfo{volume}{11}},
  \bibinfo{pages}{261} (\bibinfo{year}{1971}).

\bibitem[{\citenamefont{{Yi-Hsin Liu} et~al.}(2011)\citenamefont{{Yi-Hsin Liu},
  Drake, and Swisdack}}]{yhliu11b}
\bibinfo{author}{\bibnamefont{{Yi-Hsin Liu}}},
  \bibinfo{author}{\bibfnamefont{J.~F.} \bibnamefont{Drake}}, \bibnamefont{and}
  \bibinfo{author}{\bibfnamefont{M.}~\bibnamefont{Swisdack}},
  \bibinfo{journal}{in preparation}  (\bibinfo{year}{2011}).

\bibitem[{\citenamefont{Sonnerup et~al.}(1981)\citenamefont{Sonnerup,
  Paschmann, Papamastorakis, Sckopke, Haerendel, Bame, Asbridge, Gosling, and
  Russell}}]{sonnerup81b}
\bibinfo{author}{\bibfnamefont{B.~U.~{\"O}.} \bibnamefont{Sonnerup}},
  \bibinfo{author}{\bibfnamefont{G.}~\bibnamefont{Paschmann}},
  \bibinfo{author}{\bibfnamefont{I.}~\bibnamefont{Papamastorakis}},
  \bibinfo{author}{\bibfnamefont{N.}~\bibnamefont{Sckopke}},
  \bibinfo{author}{\bibfnamefont{G.}~\bibnamefont{Haerendel}},
  \bibinfo{author}{\bibfnamefont{S.~J.} \bibnamefont{Bame}},
  \bibinfo{author}{\bibfnamefont{J.~R.} \bibnamefont{Asbridge}},
  \bibinfo{author}{\bibfnamefont{J.~T.} \bibnamefont{Gosling}},
  \bibnamefont{and} \bibinfo{author}{\bibfnamefont{C.~T.}
  \bibnamefont{Russell}}, \bibinfo{journal}{J. Geophys. Res.}
  \textbf{\bibinfo{volume}{86}}, \bibinfo{pages}{10049} (\bibinfo{year}{1981}).

\bibitem[{\citenamefont{Gosling et~al.}(2005)\citenamefont{Gosling, Skoug,
  McComas, and Smith}}]{gosling05a}
\bibinfo{author}{\bibfnamefont{J.~T.} \bibnamefont{Gosling}},
  \bibinfo{author}{\bibfnamefont{R.~M.} \bibnamefont{Skoug}},
  \bibinfo{author}{\bibfnamefont{D.~J.} \bibnamefont{McComas}},
  \bibnamefont{and} \bibinfo{author}{\bibfnamefont{C.~W.} \bibnamefont{Smith}},
  \bibinfo{journal}{J. Geophys. Res.} \textbf{\bibinfo{volume}{110}}
  (\bibinfo{year}{2005}).

\bibitem[{\citenamefont{Krauss-Varban and Omidi}(1995)}]{krauss-varban95a}
\bibinfo{author}{\bibfnamefont{D.}~\bibnamefont{Krauss-Varban}}
  \bibnamefont{and} \bibinfo{author}{\bibfnamefont{N.}~\bibnamefont{Omidi}},
  \bibinfo{journal}{Geophys. Res. Lett.} \textbf{\bibinfo{volume}{22}},
  \bibinfo{pages}{3271} (\bibinfo{year}{1995}).

\bibitem[{\citenamefont{Hoshino et~al.}(1998)\citenamefont{Hoshino, Mukai, and
  Yamamoto}}]{hoshino98b}
\bibinfo{author}{\bibfnamefont{M.}~\bibnamefont{Hoshino}},
  \bibinfo{author}{\bibfnamefont{T.}~\bibnamefont{Mukai}}, \bibnamefont{and}
  \bibinfo{author}{\bibfnamefont{T.}~\bibnamefont{Yamamoto}},
  \bibinfo{journal}{J. Geophys. Res.} \textbf{\bibinfo{volume}{103}},
  \bibinfo{pages}{4509} (\bibinfo{year}{1998}).

\bibitem[{\citenamefont{Mjolhus}(1976)}]{mjolhus76a}
\bibinfo{author}{\bibfnamefont{E.}~\bibnamefont{Mjolhus}}, \bibinfo{journal}{J.
  Plasma. Phys.} \textbf{\bibinfo{volume}{16}}, \bibinfo{pages}{321}
  (\bibinfo{year}{1976}).

\bibitem[{\citenamefont{Spangler and Plapp}(1992)}]{spangler92a}
\bibinfo{author}{\bibfnamefont{S.~R.} \bibnamefont{Spangler}} \bibnamefont{and}
  \bibinfo{author}{\bibfnamefont{B.~B.} \bibnamefont{Plapp}},
  \bibinfo{journal}{Phys. Fluids B} \textbf{\bibinfo{volume}{4}},
  \bibinfo{pages}{3356} (\bibinfo{year}{1992}).

\bibitem[{\citenamefont{Barnes and Hollweg}(1974)}]{barnes74a}
\bibinfo{author}{\bibfnamefont{A.}~\bibnamefont{Barnes}} \bibnamefont{and}
  \bibinfo{author}{\bibfnamefont{J.~V.} \bibnamefont{Hollweg}},
  \bibinfo{journal}{J. Geophys. Res.} \textbf{\bibinfo{volume}{79}},
  \bibinfo{pages}{2302} (\bibinfo{year}{1974}).

\bibitem[{\citenamefont{Winske and Omidi}(1992)}]{winske92a}
\bibinfo{author}{\bibfnamefont{D.}~\bibnamefont{Winske}} \bibnamefont{and}
  \bibinfo{author}{\bibfnamefont{N.}~\bibnamefont{Omidi}}, \bibinfo{journal}{J.
  Geophys. Res.} \textbf{\bibinfo{volume}{97}}, \bibinfo{pages}{14779}
  (\bibinfo{year}{1992}).

\bibitem[{\citenamefont{Yin et~al.}(2007{\natexlab{b}})\citenamefont{Yin,
  Winske, Daughton, and Bowers}}]{yin07a}
\bibinfo{author}{\bibfnamefont{L.}~\bibnamefont{Yin}},
  \bibinfo{author}{\bibfnamefont{D.}~\bibnamefont{Winske}},
  \bibinfo{author}{\bibfnamefont{W.}~\bibnamefont{Daughton}}, \bibnamefont{and}
  \bibinfo{author}{\bibfnamefont{K.~J.} \bibnamefont{Bowers}},
  \bibinfo{journal}{Phys. Plasmas} \textbf{\bibinfo{volume}{14}},
  \bibinfo{eid}{062104} (\bibinfo{year}{2007}{\natexlab{b}}).

\bibitem[{\citenamefont{Davidson and Volk}(1968)}]{davidson68a}
\bibinfo{author}{\bibfnamefont{R.~C.} \bibnamefont{Davidson}} \bibnamefont{and}
  \bibinfo{author}{\bibfnamefont{H.~J.} \bibnamefont{Volk}},
  \bibinfo{journal}{Physics of Fluids} \textbf{\bibinfo{volume}{11}},
  \bibinfo{pages}{2259} (\bibinfo{year}{1968}).

\bibitem[{\citenamefont{Krauss-Varban et~al.}(1994)\citenamefont{Krauss-Varban,
  Omidi, and Quest}}]{krauss-varban94a}
\bibinfo{author}{\bibfnamefont{D.}~\bibnamefont{Krauss-Varban}},
  \bibinfo{author}{\bibfnamefont{N.}~\bibnamefont{Omidi}}, \bibnamefont{and}
  \bibinfo{author}{\bibfnamefont{K.~B.} \bibnamefont{Quest}},
  \bibinfo{journal}{J. Geophys. Res.} \textbf{\bibinfo{volume}{99}},
  \bibinfo{pages}{5987} (\bibinfo{year}{1994}).

\bibitem[{\citenamefont{Treumann}(2009)}]{treumann09a}
\bibinfo{author}{\bibfnamefont{R.~A.} \bibnamefont{Treumann}},
  \bibinfo{journal}{Astron. Astrophys. Rev.} \textbf{\bibinfo{volume}{17}},
  \bibinfo{pages}{409} (\bibinfo{year}{2009}).

\bibitem[{\citenamefont{Lucek et~al.}(2002)\citenamefont{Lucek, Horbury,
  Dunlop, Cargill, Schwartz, Balogh, Brown, Carr, Fornacon, and
  Georgescu}}]{lucek02a}
\bibinfo{author}{\bibfnamefont{E.~A.} \bibnamefont{Lucek}},
  \bibinfo{author}{\bibfnamefont{T.~S.} \bibnamefont{Horbury}},
  \bibinfo{author}{\bibfnamefont{M.~W.} \bibnamefont{Dunlop}},
  \bibinfo{author}{\bibfnamefont{P.~J.} \bibnamefont{Cargill}},
  \bibinfo{author}{\bibfnamefont{S.~I.} \bibnamefont{Schwartz}},
  \bibinfo{author}{\bibfnamefont{A.}~\bibnamefont{Balogh}},
  \bibinfo{author}{\bibfnamefont{P.}~\bibnamefont{Brown}},
  \bibinfo{author}{\bibfnamefont{C.}~\bibnamefont{Carr}},
  \bibinfo{author}{\bibfnamefont{K.~H.} \bibnamefont{Fornacon}},
  \bibnamefont{and}
  \bibinfo{author}{\bibfnamefont{E.}~\bibnamefont{Georgescu}},
  \bibinfo{journal}{Ann. Geophysicae} \textbf{\bibinfo{volume}{20}},
  \bibinfo{pages}{1699} (\bibinfo{year}{2002}).

\bibitem[{\citenamefont{Omidi and Winske}(1990)}]{omidi90a}
\bibinfo{author}{\bibfnamefont{M.}~\bibnamefont{Omidi}} \bibnamefont{and}
  \bibinfo{author}{\bibfnamefont{D.}~\bibnamefont{Winske}},
  \bibinfo{journal}{J. Geophys. Res.} \textbf{\bibinfo{volume}{95}},
  \bibinfo{pages}{2281} (\bibinfo{year}{1990}).

\bibitem[{\citenamefont{Scholer}(1993)}]{scholer93a}
\bibinfo{author}{\bibfnamefont{M.}~\bibnamefont{Scholer}}, \bibinfo{journal}{J.
  Geophys. Res.} \textbf{\bibinfo{volume}{98}}, \bibinfo{pages}{47}
  (\bibinfo{year}{1993}).

\bibitem[{\citenamefont{Bell}(1978)}]{bell78a}
\bibinfo{author}{\bibfnamefont{A.~R.} \bibnamefont{Bell}},
  \bibinfo{journal}{Mon. Not. R. astr. Soc.} \textbf{\bibinfo{volume}{182}},
  \bibinfo{pages}{147} (\bibinfo{year}{1978}).

\bibitem[{\citenamefont{Lee}(1982)}]{lee82a}
\bibinfo{author}{\bibfnamefont{M.~A.} \bibnamefont{Lee}}, \bibinfo{journal}{J.
  Geophys. Res.} \textbf{\bibinfo{volume}{87}}, \bibinfo{pages}{5063}
  (\bibinfo{year}{1982}).

\bibitem[{\citenamefont{Fisk et~al.}(2006)\citenamefont{Fisk, Gloeckler, and
  Zurbuchen}}]{fisk06a}
\bibinfo{author}{\bibfnamefont{L.~A.} \bibnamefont{Fisk}},
  \bibinfo{author}{\bibfnamefont{G.}~\bibnamefont{Gloeckler}},
  \bibnamefont{and} \bibinfo{author}{\bibfnamefont{T.~H.}
  \bibnamefont{Zurbuchen}}, \bibinfo{journal}{Astrophys. J.}
  \textbf{\bibinfo{volume}{644}}, \bibinfo{pages}{631} (\bibinfo{year}{2006}).

\bibitem[{\citenamefont{Decker}(1988)}]{decker88a}
\bibinfo{author}{\bibfnamefont{R.~B.} \bibnamefont{Decker}},
  \bibinfo{journal}{Space Science Reviews} \textbf{\bibinfo{volume}{48}},
  \bibinfo{pages}{195} (\bibinfo{year}{1988}).

\bibitem[{\citenamefont{Sagdeev}(1966)}]{sagdeev66a}
\bibinfo{author}{\bibfnamefont{R.~Z.} \bibnamefont{Sagdeev}},
  \bibinfo{journal}{Reviews of Plasma Physics} \textbf{\bibinfo{volume}{4}},
  \bibinfo{pages}{23} (\bibinfo{year}{1966}).

\end{thebibliography}

\end{document}